\documentclass[twocolumn,showpacs,preprintnumbers,amsmath,amssymb]{revtex4}
\usepackage{floatflt}
\usepackage{amsfonts}
\usepackage{amssymb}
\usepackage{psfig}
\usepackage{mathrsfs}
\newcommand{\beq}{\begin{equation}}
\newcommand{\eeq}{\end{equation}}
\newcommand{\beqn}{\begin{eqnarray}}
\newcommand{\eeqn}{\end{eqnarray}}
\newcommand{\bea}[1]{\beq\begin{array}{#1}}
\newcommand{\eea}{\end{array}\eeq}

\newcommand{\Tr}[1]{\;{1\over #1}\mathop{\rm Tr}}
\newcommand{\tr}{\mathop{\rm Tr}}

\newcommand{\im}[1]{\mathrm{Im}\,#1}

\newcommand{\Pexp}{\mbox{P}\!\exp}

\newcommand{\ket}[1]{|\,#1\,\rangle}
\newcommand{\bra}[1]{\langle\,#1\,|}
\newcommand{\braket}[2]{\langle\,#1\,|\,#2\,\rangle}

\newcommand{\diff}{\partial}

\begin{document}
\preprint{ITEP-LAT/2005-07}

\title{$\langle A^2 \rangle$ Condensate, Bianchi Identities and Chromomagnetic\\
       Fields Degeneracy in SU(2) YM Theory}

\author{F.V.~Gubarev}
 \email{gubarev@itep.ru}
\author{S.M.~Morozov}
 \email{smoroz@itep.ru}
 \altaffiliation[Also at ]{Moscow Institute of Physics and Technology,
                           Dolgoprudniy, Moscow region, Russia}
\affiliation{Institute of Theoretical and  Experimental Physics,
             B.Cheremushkinskaya 25, Moscow, 117259, Russia}

\begin{abstract}
\noindent
We consider the non-Abelian Bianchi identities in SU(2) pure Yang-Mills theory in D=3,4 focusing
on the possibility of their violation and the significance of the chromomagnetic fields degeneracy points.
We show that the recently proposed non-Abelian Stokes theorem allows to formulate the Bianchi
identities in terms of the physical fluxes and their relative color orientations.
Then the violation of Bianchi identities becomes a well defined concept ultimately related to the degeneracy points. 
The locality and gauge invariance of our approach allows to study the problem numerically.
We present evidences that in D=4 the suppression of the Bianchi identities violation is likely to destroy
confinement while the removal of the degeneracy points drives the theory to the topologically non-trivial sector.
However, confronting the results obtained in three and four dimensions we argue that it is the mass
dimension two condensate $\langle A^2_{min} \rangle$ which probably explains our findings.
\end{abstract}


\maketitle
\section{Introduction}
\noindent
Gauge theories are usually formulated in terms of the gauge potentials $A^a_\mu$
taking values in the Lie algebra of the corresponding gauge group. Provided that
the gauge coupling is small this description is indeed adequate and provides
local functionally independent coordinates on the configuration space.
However, in the strongly coupled gauge theories the potentials themselves 
obtain a separate physical meaning. Here we mean the non-perturbative
dimension 2 condensate $\langle A^2_{min}\rangle$ introduced in~\cite{Gubarev:2000eu,Gubarev:2000nz},
which received a particular attention in recent years (see, e.g. Ref.~\cite{Dudal:2005na} for review
and further references).

The original motivation of this work was the analysis of various possible contributions
to the  $\langle A^2_{min}\rangle$ condensate. Note that the central point of Ref.~\cite{Gubarev:2000eu}
was in fact the consideration of the Abelian Bianchi identities and their ultimate relation
to $\langle A^2_{min}\rangle$. As far as the Abelian theory is concerned the non-triviality
of $\langle A^2_{min}\rangle$ condensate is essentially equivalent to the Bianchi identities
violation. Therefore in the non-Abelian case it seems natural to start
from the corresponding Bianchi identities and investigate their role in the $\langle A^2_{min}\rangle$ condensate formation.
However, the literature on the subject turns out to be scarce. In particular, as is well known from the Abelian models
the rigorous treatment of the Bianchi identities requires the non-perturbative (say, lattice)
regularization. But we were unable to find papers devoted to this problem in the non-Abelian case.

On the other hand, the investigation of the non-Abelian Bianchi identities is important on its own right.
Without mentioning all the aspects of the problem, let us note that the $\langle A^2_{min}\rangle$
condensate is certainly connected with the non-Abelian Bianchi identities.
Moreover, it was emphasized in Refs.~\cite{Gubarev:2002kk,Zakharov:2003nm,Zakharov:2005md}
that the Bianchi identities and the possibility of their violation are ultimately related to the confinement problem. 
Then the logic suggests to consider whether the $\langle A^2_{min}\rangle$ condensate is
relevant for confinement as well, the question which was discussed in~\cite{Gubarev:2000nz,Zakharov:2003nm}
(see also~\cite{Suzuki:2004dw}). Therefore we see that all these problems are in fact
indispensable from each other and cannot be considered separately.
We decided to focus on the Bianchi identities in this paper;
the connection with the quantities like $\langle A^2_{min}\rangle$ is discussed in the due course.
Throughout the paper we work with Euclidean three and four dimensional SU(2) gluodynamics
keeping in mind the lattice regularization of the theory, although we nowhere relay
exclusively on the lattice. The paper is reasonably self-contained, the results which we're using
are briefly reviewed. Note that the similar in spirit but in no way identical treatment could
be found in Refs.~\cite{Gubarev:2002kt,Gubarev:2002ku}.

The primary tool of our analysis is the non-Abelian Stokes theorem~\cite{Gubarev:2003ij}
derived recently by one of us. The advantage is that it allows to work directly in terms of
the gauge invariant quantities like magnitudes of the elementary fluxes and their relative orientations.
As might be expected the non-Abelian Bianchi identities could be reduced
to the application of the above theorem to the infinitesimal closed surfaces. However,
in this case the non-Abelian Stokes theorem not necessary gives zero, the answer, in fact, is
proportional to the integer number. Since every step in the derivation is gauge invariant
this integer is gauge invariant as well and in the continuum language corresponds
to the non-Abelian Bianchi identities violation.

The non-Abelian nature of the theory manifests itself in the complicated geometry underlying
the Bianchi identities. We consider all these questions in detail and show that the careful
but purely geometrical treatment leads to the consideration of the special degenerate points
in the configuration space at which a particular determinants constructed from chromoelectric
and chromomagnetic fields vanish. Finally we show that the investigation of the non-Abelian Bianchi
identities is indispensable from the study of these degenerate points.
Therefore the framework outlined above naturally extends to include the degeneracy points,
the relevance of which both for confinement and chiral symmetry breaking was discussed
in Refs.~\cite{Zakharov:2005md,Zakharov:2004jh}.

The locality and gauge invariance of our construction allow us to study the problem numerically.
We investigate the effects due to the Bianchi identities violation and the degenerate points
in the numerical simulations. As might be {\it a priori} expected the suppression of
the degenerate points always lead to the violation of the reflection positivity.
Moreover, in D=4 one could easily pin-point the origin of the reflection positivity
violation: it is caused by rapidly rising global topological charge.  Thus in D=4
the suppression of the degenerate points shifts the vacuum to the non-trivial topological sector.

As far as the Bianchi identities are concerned the results depend crucially on the space-time dimensionality.
In D=3 the suppression of the Bianchi identities violation does not change the theory in any notable way.
However, in D=4 the effect is different: it seems that
the suppression of the Bianchi identities violation is likely to destroy confinement 
while other measured characteristics of the theory remain qualitatively unchanged.
At least this is so for the lattices and coupling constants we have considered.
Note that the problem still requires a careful numerical investigation, in particular,
we had not studied yet the volume dependence of our results. The corresponding analysis
will be published elsewhere.

Finally we argue that it would be misleading to interpret our results as the statement that
confinement is caused by the Bianchi identities violation. Confronting the results obtained
in three and four dimensions we show that it is the
$\langle A^2_{min}\rangle$ condensate which is probably relevant for confinement. Although
the argumentation is not rigorous it seems to be the only one which matches our findings.

\section{Formulation of the Problem}
\label{bianchi-intro}
\noindent
The primary object of our investigation is the Bianchi identities for
SU(2) gauge fields in four space-time dimensions. Thus we will analyze the equations
\bea{c}
\label{bianchi-D4}
\diff_\mu \tilde{F}^a_{\mu\nu} + \varepsilon^{abc} A^b_\mu \tilde{F}^c_{\mu\nu} = 0\,, \\
\\
\tilde{F}^a_{\mu\nu} = \frac{1}{2}\, \varepsilon_{\mu\nu\lambda\rho}\, F^a_{\lambda\rho}
\qquad [D=4]\,,
\eea
having in mind eventually Euclidean lattice regularization of SU(2) pure Yang-Mills theory.
Here $F^a_{\mu\nu}$ is the conventional continuum field-strength tensor
\beq
\label{field-strength}
F^a_{\mu\nu} = \diff_\mu A^a_\nu - \diff_\nu A^a_\mu + \varepsilon^{abc} A^b_\mu A^c_\nu\,,
\eeq
Greek and Latin indexes run through $0,..,3$ and $1,...,3$ respectively.
Our treatment also applies in three dimensions where Bianchi identities are as follows
\bea{c}
\label{bianchi-D3}
\diff_i \, B^a_i \,+\, \varepsilon^{abc} A^b_i B^c_i ~=~ 0\,, \\
\\
B^a_i = \frac{1}{2}\,\varepsilon_{ijk}\,F^a_{jk} \qquad [D=3]\,.
\eea
However, it turns out that the three-dimensional case is physically quite different
from $D=4$ and we'll comment on that in the due course.

In this section we give qualitative continuum arguments which show that
at least at some points in the configuration space the Bianchi identities
(\ref{bianchi-D4}), (\ref{bianchi-D3}) should be considered with care.

\subsection{Chromomagnetic fields degeneracy}
\label{degeneracy}
\noindent
It has been known for a long time that in non-Abelian gauge theories two or more gauge inequivalent
potentials could produce the same field strength~\cite{Wu:1975es}.
This phenomenon, known as Wu-Yang ambiguity, had received great attention in the past
(see, e.g. \cite{Gu:1977sy,Deser:1976wj,Bollini:1979ah})
and it was noted long ago  \cite{Roskies:1976ux,Calvo:1977tx,Halpern:1977fw}
that in D=4 the Bianchi identities constitute an algebraic
obstruction for the ambiguity to exist. Namely, for given chromoelectric
$E^a_i = F^a_{0i}$ and chromomagnetic $B^a_i = 1/2\,\varepsilon_{ijk}\,F^a_{jk}$ 
fields Eq.(\ref{bianchi-D4}) is a linear algebraic system of 
12 equations for 12 unknown $A^a_\mu$. Therefore away from the set of points where
the matrix $T^{ab}_{\mu\nu} ~=~ \varepsilon^{abc}\, \tilde{F}^c_{\mu\nu}$ degenerates
\beq
\label{degenerate}
\mathrm{det}\, T ~=~  0
\eeq
Bianchi identities allow to express the gauge potentials as local single-valued function
of $E^a_i$ and $B^a_i$. On the other hand, there is no physical principle
or symmetry which could keep the sign of $\mathrm{det}\,T$ fixed. Indeed, in the weak coupling
perturbation theory the sign of $\mathrm{det}\,T$ changes wildly
and therefore the degeneracy of chromomagnetic fields, Eq.~(\ref{degenerate}), is, in a sense, generic.
Note that the situation is quite different in D=3 since Eq.~(\ref{bianchi-D3}) formally
constitutes 3 equations for 9 unknown variables.
Therefore in three dimensions the Bianchi identities do not constrain
the gauge potentials at all and the Wu-Yang ambiguity problem is much more severe
(see, e.g. Refs.~\cite{Freedman:1994dv,Majumdar:1998cc} for discussion).
Unfortunately, we are not aware of any conclusive considerations of the degenerate points
(\ref{degenerate}) in the literature. It is true that Eq.~(\ref{degenerate}) by itself
is known for a long time~\cite{Halpern:1977fw,Halpern:1977ia,Deser:1976iy}
but most of the analysis performed so far considered it
in the context of dual formulation of gluodynamics \cite{Lunev,Ganor:1995em,Bauer:1994hj,Haagensen:1994sy,Diakonov:2001xg}
from which the information about original Yang-Mills fields is hard to extract.
Ref.~\cite{Freedman:1993mu} seems to be the only exception where it was argued that physical
wave functionals should vanish at the points of degeneracy. We will see below that equations similar to
(\ref{degenerate}) arise naturally in the construction of the Bianchi identities.
Moreover, the points of degeneracy seem to be relevant for gauge fields dynamics.

What we have said so far is in accordance with general expectation that in the non-Abelian gauge
theories there is no unique way to express $A^a_\mu$ in terms of the corresponding field strength
(apart from the usual gauge ambiguity, of course). At this point one could 
give an example of special gauges (complete axial, coordinate, contour gauges,
see~\cite{Shevchenko:1998uw,DiGiacomo:2000va}
for review), in which the gauge potentials are always explicit single-valued functions of the field
strength. Is there any contradiction? Although this question is not directly related to our
work, we note that all the gauges mentioned above are consistent only if
Bianchi identities (\ref{bianchi-D4}), (\ref{bianchi-D3}) are satisfied identically~\cite{Halpern:1978ik}.
In particular, in the Abelian case one notices~\cite{Shevchenko:1998uw} that the presence
of elementary magnetic charges forces the potentials in contour gauge to depend upon
the arbitrary contour prescription. Of course, this is a manifestation of famous
Wu-Yang ambiguity which in this case certainly arises because point-like monopoles violate
the Bianchi identities. We conclude therefore that the possibility of Bianchi
identities violation should not be excluded {\it a priori}. Moreover, the very existence
of Wu-Yang ambiguous potentials hints on the violation of (\ref{bianchi-D4}), (\ref{bianchi-D3}).

\subsection{Bianchi identities violation}
\label{violation-intro}
\noindent
The possibility that the r.h.s. of Eqs.~(\ref{bianchi-D4}), (\ref{bianchi-D3})
might be non-zero was considered long ago (see, e.g.~\cite{Halpern:1978ik}),
but as far as we know this approach had never been actively developed. 
This is mostly because the study of Bianchi identities
violation requires a particular regularization, which should correctly respect the global
structure of the gauge group. It turns out that for our purposes the lattice formulation
is distinguished (see Refs.~\cite{Chernodub:2000wk,Chernodub:2000rg} for discussion).
Therefore consider the basic SU(2) gauge theory observable, which is also the fundamental
object on the lattice, the Wilson loop in spin $1/2$ representation
\bea{c}
\label{wilson-intro}
W(C,x_0) ~=~ \Pexp \, i\sigma^a \oint_{C(x_0)} A^a_\mu \, dx^\mu\,,\\
\\
W(C) ~=~ \Tr{2} \, W(C,x_0)\,.
\eea
Here $\sigma^a$ are the Pauli matrices, $C$ is some closed contour with marked point
$x_0 \in C$ from which the path ordered integral starts and $\mathrm{P}$-ordering
is defined from left to the right. Note the unusual normalization of SU(2) generators
which we take for future convenience. By definition the operator
$W(C,x_0)$ measures the non-Abelian flux $\Phi(C,x_0)$ penetrating the contour
\bea{c}
\label{flux-intro}
W(C,x_0) = e^{ i \sigma^a\,\Phi^a(C,x_0)}\,,
\qquad W(C) = \cos \,\Phi(C)\,, \\
\\
\Phi(C) = \sqrt{ \Phi^a(C,x_0) \, \Phi^a(C,x_0)}\,,
\eea
where the flux \footnote{
Strictly speaking this is a flux magnitude, however we often use the term ``flux''
since it is clear from context which quantity is meant.
} $\Phi(C)$ is gauge invariant and does not depend on $x_0$. Eq.~(\ref{wilson-intro})
will be thoroughly analyzed later, but now we note that
the physically observable flux is always bounded $0 < \Phi(C) < \pi$
due to periodicity (compactness) of the gauge action.
Moreover, there exist no physically meaningful experiment
which could distinguish the fluxes $\Phi(C)$ and $\Phi(C) + 2\pi$ and
this observation applies equally well to the infinitesimal contours
which constitute the lattice definition of the field strength.
On the other hand, there is no trace whatsoever of the gauge action compactness
in the continuum expression (\ref{field-strength}).
In this respect the SU(2) gluodynamics is similar to the compact
U(1) gauge model~\cite{Polyakov:1975rs} (see Ref.~\cite{Peskin:1977kp} for review).
In fact, some consequences of the compactness of the non-Abelian gauge theories 
were already discussed in the past~\cite{Orland:1982fv}.
Note however that we are not saying that singular fluxes are important in the continuum
limit of lattice formulation. After all this is a dynamical question which cannot be
studied with simple arguments above. Rather we would like to point out that
the very definition of $F^a_{\mu\nu}$ on the lattice is {\it a priori} different from
the continuum one (\ref{field-strength}) and therefore the validity of
(\ref{bianchi-D4}), (\ref{bianchi-D3}) in the lattice context should be considered anew.
We stress that our arguments are purely kinematical
and follow directly from the gauge invariance along.
Whether or not the violation of Bianchi identities is physically relevant 
is a dynamical issue which we investigate (at least partially) later on.

To conclude we note that nowadays there exist both theoretical
arguments~\cite{Gubarev:2002kk,Zakharov:2003nm,Zakharov:2005md}
and the experimental lattice data~\cite{Skala} which favor the non-vanishing
r.h.s. of Eqs.~(\ref{bianchi-D4}), (\ref{bianchi-D3}) in the continuum limit
of lattice gauge models. Although the approaches of these papers are quite different, 
the conclusion is essentially the same: the non-Abelian Bianchi identities are indeed
violated in the scaling (continuum) limit and this fact is related to the problem of confinement.

\section{Lattice Bianchi Identities}
\subsection{Preliminaries}
\label{program}
\noindent
In this section we briefly summarize what has been known so far about the non-Abelian Bianchi identities
on the lattice and comment on the strategy we employ in this paper. Surprisingly enough
the literature on the subject seems to be very scarce (contrary to the Abelian case which
we do not consider however) and the most relevant for
our discussion references are~\cite{Kiskis:1982ty,Batrouni:1981ri,Batrouni:1984rb}
(see also~\cite{Orland:1982fv}). Historically, the Bianchi identities 
explicitly appeared first in the context of plaquette (field-strength) formulation
of lattice QCD~\cite{Batrouni:1981ri,Batrouni:1984rb}. In particular, it was noted
that the strong coupling expansion can be obtained as an expansion towards restoring
the lattice Bianchi identities.

\begin{figure}
\centerline{\psfig{file=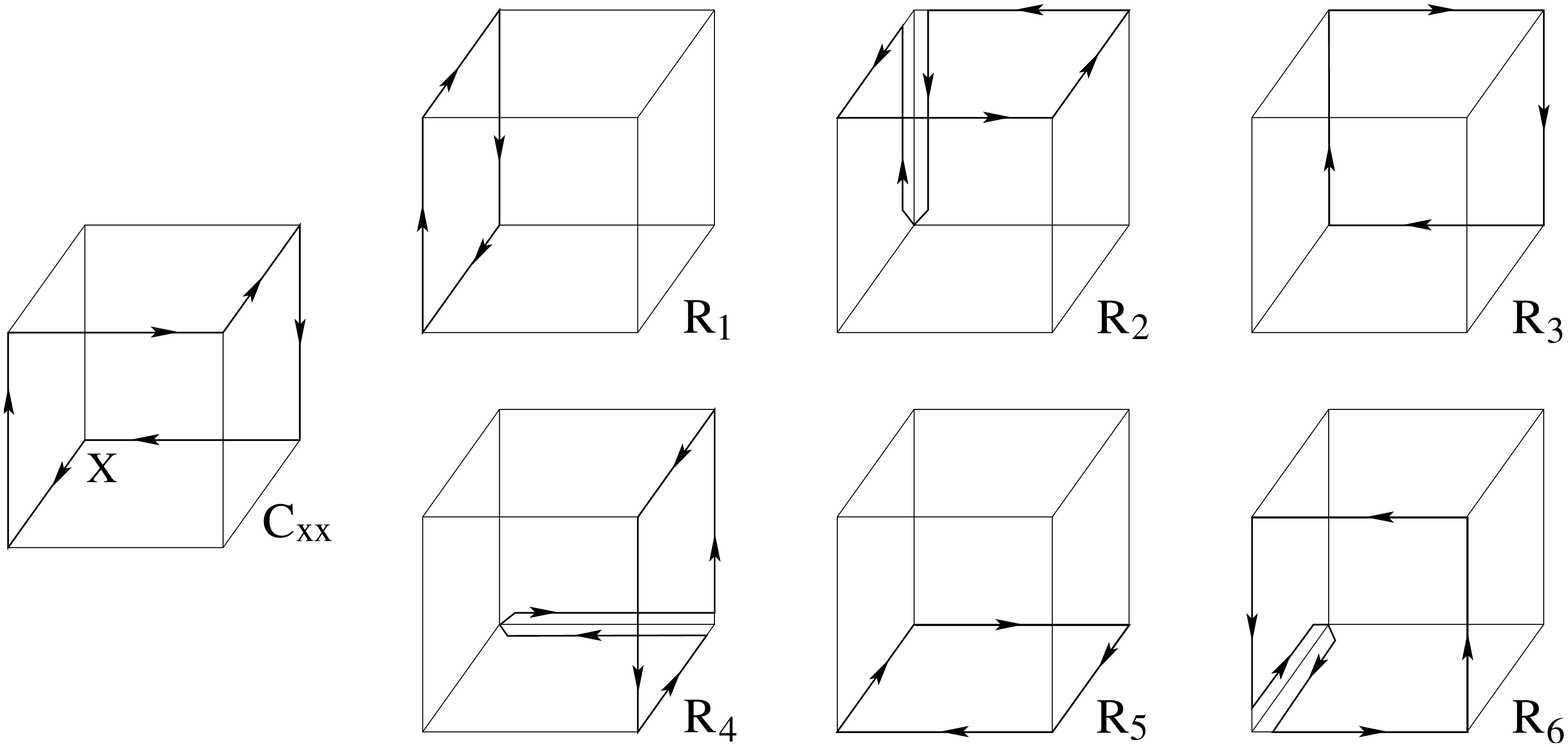,width=0.3\textwidth,silent=}}
\caption{Graphical representation of lattice Bianchi identities.}
\label{fig1}
\end{figure}

It turns out that the formulation of Ref.~\cite{Kiskis:1982ty} is the most appropriate
for our purposes. Essentially it consists in the observation
that any lattice gauge field configuration could be interpreted as a homomorphism from
the lattice edge path group into the gauge group (see Ref.~\cite{Dubrovin} for definitions).
It follows form the definition of homomorphic mapping that
\beq
\label{kiskis}
U( C_{xy} \circ C^{-1}_{xy} ) = U( C_{xy}) U^{-1}(C_{xy} ) = 1\,,
\eeq
where $C_{xy}$ is arbitrary path connecting the points $x$ and $y$ and the composite path
$C_{xy} \circ C^{-1}_{xy}$ is usually referred to as null-homotopic. In fact, Eq.~(\ref{kiskis})
looks rather obvious for everyone familiar with lattice formulation. However, the assertion
of Ref.~\cite{Kiskis:1982ty} is that Eq.~(\ref{kiskis}) constitutes the most general
form of lattice Bianchi identities and indeed just that: an identity. 
Note that Eq.~(\ref{kiskis}) looks quite different from what is expected in the continuum.
To establish the relation between (\ref{kiskis}) and (\ref{bianchi-D4}),(\ref{bianchi-D3}) consider
the path $C_{xx}$ shown on Figure~\ref{fig1}.
It follows trivially that the equality $U(C_{xx} \circ C^{-1}_{xx}) = 1$ is equivalent to
\beq
\label{kiskis2}
U(R_1)\, U(R_2)\, U(R_3)\, U(R_4)\, U(R_5)\, U(R_6) ~=~ 1\,,
\eeq
which in the naive continuum limit reduces to the conventional Bianchi identities
(\ref{bianchi-D4}), (\ref{bianchi-D3}). Moreover, Eq.~(\ref{kiskis2}) is
the particular case of the so called operator non-Abelian Stokes
theorem~\cite{Arefeva:dp}-\cite{Simonov:xs} (see, e.g.~\cite{Shevchenko:1998uw} for review)
which allows to represent (rather formally though)
the path ordered exponent as the surface ordered integral
\beq
\label{NAST-operator}
\Pexp\,i\int\limits_{C=\delta S_C} A_\mu\, dx^\mu  =
\mbox{P}_S \exp\frac{i}{2}\int\limits_{S_C} \mathscr{F}_{\mu\nu} \, d^2 \sigma^{\mu\nu}\,,
\eeq
where $\mathscr{F}$ is non-local covariantly transformed field-strength the concrete form of
which is not important for what follows. The surface
$S_C$ is arbitrary and consistency requires the representation (\ref{NAST-operator})
to be independent on $S_C$ as long as $\delta S_C = C$. In particular,
the r.h.s. of Eq.~(\ref{NAST-operator})
being applied to closed surface $S_0$, $\delta S_0 = 0$, should always give the identity
\beq
\label{bianchi-NAST}
\mbox{P}_S \exp\,\frac{i}{2}\int_{S_0,\delta S_0 = 0}
\mathscr{F}_{\mu\nu} \, d^2 \sigma^{\mu\nu} ~=~ 1 \,.
\eeq
In fact, Eq.~(\ref{kiskis2}) is the special case of (\ref{bianchi-NAST}) in which
$S_0$ is the boundary of elementary lattice cube. Therefore, it seems to be
legitimate to formulate the non-Abelian Bianchi identities
as the requirement of surface independence of the non-Abelian Stokes theorem.

Eqs.~(\ref{kiskis})-(\ref{bianchi-NAST}) are the starting point of our considerations below.
However, before going into details let us comment a bit on our strategy. We note first that
the identity on the r.h.s. of Eqs.~(\ref{kiskis2}), (\ref{bianchi-NAST}) could in general
be written as
\beq
\label{magnetic-charge}
1 ~=~ e^{\,i\, \vec{\sigma}\vec{n}\cdot 2\pi q}\,,
\qquad \vec{n}^2 = 1\,,
\qquad q\, \in \,Z\,.
\eeq
The color direction $\vec{n}$ is gauge variant and will not concern us here.
Suppose that we are able to give an unambiguous gauge invariant meaning to the integer $q$ and
that it is non-zero for some $S_0$ in given gauge background.
Then this would certainly mean that there is a point \footnote{
We take D=3 for simplicity, the modifications for D=4 are obvious.
} somewhere inside $S_0$ at which the continuum Bianchi identities are violated.
Here the argumentation is essentially the same as in well known Abelian case.
So the problem is to make sense of $q$ which should be well defined and gauge invariant.
{}From now on we refer to the integer $q$ as the ``magnetic charge'' whatever it is.
In particular, neither charge conservation nor any other usual properties of the magnetic
charge are assumed.
Secondly, Eqs.~(\ref{kiskis2}),(\ref{bianchi-NAST}) are not quite suitable to analyze the
Bianchi identities. This is precisely because neither (\ref{kiskis2}) nor (\ref{bianchi-NAST})
make, in fact, no direct reference to the non-Abelian field strength.
And this is in sharp contrast with the Abelian theory in which the Bianchi identities even on the lattice explicitly
refer to physical fluxes. It turns out that the solution of the second problem simultaneously solves the
first, namely, the non-Abelian Stokes theorem being expressed in terms of the physical field strength
provides the definition of $q$ which we are looking for.

\subsection{Chromomagnetic Fields on the Lattice}
\label{chromomagnetic-fields}
\noindent
The distinguished feature of the lattice regularization is that the gauge theory is formulated
in terms of the Wilson loops along and strictly speaking the lattice does not
need to introduce the notion of the field strength. Chromomagnetic fields appear
only in the limit of vanishing lattice spacing, overwise one should rather think in terms
of the non-Abelian fluxes which are defined by Eqs.~(\ref{wilson-intro}), (\ref{flux-intro}).
Therefore consider the Wilson loop
\bea{c}
\label{wilson}
W(C,t) = \Pexp\,i\int^{T+t}_t A(\tau)\,d\tau
= e^{\,i\vec{\sigma} \, \vec{n}(C,t)\cdot \Phi(C)}\,, \\
\\
A(\tau) = \sigma^a \, A^a_\mu(x) \, \dot{x}^\mu(\tau)\,, \qquad \vec{n}^2(C,t) = 1\,,\\
\\
W(C) = \Tr{2} W(C,t) = \cos \,\Phi(C)\,,
\eea
defined for some closed contour $C=\{\,x(t),\,0\le t\le T,\,x(0)=x(T)\}$
(our presentation is similar but not identical to that of Ref.~\cite{Gubarev:2003ij},
see also Ref.~\cite{Gubarev:2000qg}).
We assume that $W(C) \ne \pm 1$ and then it is convenient to parametrize
the Wilson loop in terms of the flux magnitude
$\Phi(C)\in (0\,;\,\pi)$ and the instantaneous flux direction in color space $\vec{n}(C,t)$
which explicitly depends on $t$. It is clear that $\Phi(C)$ is gauge invariant while $\vec{n}(C,t)$
rotates as three-dimensional vector under the gauge transformations at point $x(t)$.
Consider now another contour $C'$ which touches (or intersects) $C$ at point $x(t_0)=x'(t'_0)$.
Evidently, while both $\vec{n}(C,t_0)$ and $\vec{n}(C',t'_0)$ are gauge variant their relative orientation 
(angle in between) is gauge independent. Moreover, the construction
could be iterated: for any number of contours intersecting at one point the relative
orientation of instantaneous fluxes at that point is gauge invariant.
It is amusing to note that the relative orientation of elementary fluxes received almost no
attention in the past. While the magnitude of various fluxes had been discussed and measured 
in various circumstances (see, e.g. Ref.~\cite{Stephenson:1998dd} and references therein), it seems that
only Refs.~\cite{Kiskis:1983fa,Kiskis:1984ru} studied their relative orientations.

Consider next the behavior of the flux parametrized by Eq.~(\ref{wilson})
under the change of contour orientation. Physically one expects that the total flux
should change sign when contour is followed in the opposite direction
\beq
\label{orientation-change-flux}
\Phi^a(C^{-1},t) = \Phi(C^{-1}) \, n^a(C^{-1},t) = - \Phi^a(C,t)\,.
\eeq
The parametrization (\ref{wilson}) respects the intuition and indeed the flux direction
changes sign while the flux magnitude is orientation independent
\beq
\label{orientation-change}
\vec{n}(C^{-1},t) = - \vec{n}(C,t)\,, \qquad \Phi(C^{-1}) = \Phi(C)\,.
\eeq

Here we come to the important point concerning the determination of physical field
strength from the infinitesimal fluxes. Suppose that we measure twice the elementary
flux, first with an oriented area element $\delta\sigma^{\mu\nu}$ and then with
reversed orientation $\delta\sigma^{\nu\mu} = - \delta\sigma^{\mu\nu}$.
Evidently, the corresponding Wilson loops are conjugated to each other
\beq
\label{wilson-conj}
W(\delta\sigma^{\mu\nu}) ~=~ W^\dagger(\delta\sigma^{\nu\mu})\,.
\eeq
On the other hand, the expansion in powers of lattice spacing $a$ reads
$$
W(\delta\sigma^{\mu\nu}) ~=~ 1 ~+~ a^2\,i\vec{\sigma} \, \vec{F}_{\mu\nu}\delta\sigma^{\mu\nu}
~+~ O(a^4)\,,
$$
$$
W(\delta\sigma^{\nu\mu}) ~=~ 1 ~+~ a^2\, i \vec{\sigma} \, \vec{F}_{\nu\mu}\delta\sigma^{\nu\mu} ~+~ O(a^4) ~=~
W(\delta\sigma^{\mu\nu})\,.
$$
and disagrees with (\ref{wilson-conj}). This simple exercise which applies equally in the Abelian case
shows that the lattice area element $dx^\mu dx^\nu$ is in fact unoriented
$dx^\mu dx^\nu = dx^\nu dx^\mu$ contrary to the usual continuum relation
$\delta\sigma^{\mu\nu} = dx^\mu \wedge dx^\nu = - dx^\nu \wedge dx^\mu$. Therefore
in order to define the field strength on the lattice a canonical orientation
of all elementary squares (plaquettes) should be fixed first. Overwise the field strength will
suffer from sign ambiguity on different plaquettes. In fact, the canonical ordering is well known
in lattice community and the conventional agreement is to consider $\delta\sigma^{\mu\nu}$
with $\mu < \nu$ only. However, the orientation conventions are crucial for 
the interpretation of lattice equations below in the continuum terms. {}From now on
we always assume that the infinitesimal fluxes are constructed with canonically
oriented plaquettes.

It is convenient to generalize the representation (\ref{wilson}) in order to gain
a simple physical interpretation. Namely, it is natural to describe the instantaneous flux
direction by means of fictitious (iso-)spin $1/2$ particle living on the contour.
The spinor wave function is given by two-component normalized complex quantity
\beq
\bra{z} = [\,z_1\,;\,z_2\,]\,,
\qquad \braket{z}{z} = |z_1|^2 + |z_2|^2 = 1\,,
\eeq
which is bra-vector in accordance with our left to the right $\mathrm{P}$-ordering convention.
The defining equation for the Wilson loop becomes the Schr\"odinger equation for spinor
\bea{c}
\label{evolution}
\bra{z(t)}\,( i \overleftarrow{\diff}_t + A ) = 0\,, \\
\\
\bra{z(t)} = \bra{z(0)} \cdot \Pexp\,i \int^t_0 A(\tau)\,d\tau\,.
\eea
Therefore the Wilson loop (\ref{wilson}) is the quantum mechanical evolution operator for
spin degrees of freedom. As is usual in quantum mechanics the state vectors could be
arbitrary rephased
\beq
\label{rephasing}
\bra{z(t)} \, \to \, e^{i\,\theta(t)}\,\bra{z(t)}\,.
\eeq
The particular choices $\im z_1 = 0$, $\im z_2 = 0$ lead to well known families of (anti)holomorphic
spin coherent states~\cite{Perelomov} (see, e.g.~\cite{Zhang:1999is} for review).
Following the quantum mechanical analogy~\cite{Aharonov:1987gg,bhandari} one could argue
that the eigenstate of the evolution operator $W(C,0)$
\beq
\label{eigen}
\bra{z(0)} \, W(C,0) ~=~ e^{ i \, \Phi(C) } \, \bra{z(0)}\,,
\eeq
is of special importance and is usually referred to as cyclic state.
In particular, the state $\bra{z(0)}$ being the eigenstate
of $W(C,0)$ at $t=0$ remains the eigenstate of $W(C,t)$ during the evolution (\ref{evolution}).
It follows immediately that the cyclic state (\ref{eigen}) is best suited to describe the instantaneous
flux direction. Indeed, it is a matter of one-line calculation to show that
$n^a(C,t) ~=~ \bra{z(t)} \,\sigma^a\,\ket{z(t)}$.
In other words,  the flux direction $\vec{n}(C,t)$ and the ratio $z_2(t)/z_1(t)$ of cyclic state
components are related to each other by standard stereographic projection.
In particular, the flux magnitude is given by
\beq
\label{flux}
\Phi(C) ~=~ \mathrm{arg}[\,\bra{z(t)}W(C,t)\ket{z(t)}\,]
\eeq
and is t-independent. Moreover, if contour $C$ is subdivided into $N$ segments then
\beq
\label{flux2}
\! \Phi(C) = \mathrm{arg} \prod\limits_{k=0}^{N-1}
\bra{z(t_k)}\Pexp\, i\!\!\!\int\limits^{t_{k+1}}_{t_k} \!\!\! A(\tau)\,d\tau \,\ket{z(t_{k+1})} \,,
\eeq
where the identification $t_0 = t_N$ is assumed. As far as the relative orientation of fluxes
is concerned it is tempting to consider the quantities like $\mathrm{arg} \braket{z}{\zeta}$. However,
it is not invariant under~(\ref{rephasing}) because $\bra{z}$ and $\bra{\zeta}$ 
could be rephased independently. Nevertheless, the equations we will get do indeed
include the products like $\braket{z}{\zeta}$ yet respecting the U(1) invariance~(\ref{rephasing}).

It remains only to consider the multivaluedness of the cyclic state defining equation (\ref{eigen}).
Indeed, there exist two solutions of Eq.~(\ref{eigen}) while we discussed only one of them.
The second eigenstate is obtainable from the first one by substitution
\beq
\label{negate}
z_2\,\to\, z_1^* \qquad z_1\,\to\, -z_2^*\,.
\eeq
It is clear that Eq.~(\ref{eigen}) corresponds to the ``spin-up'' wave function for
which the spin is aligned with the magnetic field, while the second eigenstate (\ref{negate}) is the ``spin-down''
state which has spin anti-aligned. Our original goal was to describe the direction
of instantaneous flux and therefore the anti-aligned state should be discarded since
it corresponds to the inverted flux direction. Note also that the flux magnitude $\Phi(C)$
is positive by definition but with anti-aligned state we get $\Phi(C) < 0$.
We conclude therefore that for given contour orientation there is no ambiguity in Eq.~(\ref{eigen})
and the appropriate family of cyclic states $\bra{z(t)}$ is uniquely defined.
The second ``spin-down'' eigenstate describes the flux direction for inverted contour orientation and
therefore Eq.~(\ref{negate}) corresponds to the time reversal operation for spinors in quantum mechanics.

The above considerations apply immediately on the lattice. The only difference with the continuum
is that the gauge potentials are unknown, we have only the parallel transporters along the elementary  links.
But this is actually enough: the Wilson loop is constructed by direct matrix multiplication 
and then Eq.~(\ref{eigen}) applies literally. The instantaneous flux direction 
is determined via (\ref{eigen}) or (\ref{evolution}) at lattice sites passed by Wilson loop.
The flux magnitude is given by Eqs.~(\ref{flux}), (\ref{flux2}).

To summarize, every Wilson loop \footnote{
The reservation $W(C)\ne\pm 1$ corresponds to measure zero set of configurations and hence
irrelevant.
} $W(C,t)$ is characterized by the magnitude of the flux
$\Phi(C)$ and the instantaneous flux direction $\vec{n}(C,t)$, $\vec{n}^2 = 1$ which varies along
the contour and is reversed on changing contour orientation.
The quantum mechanical language is adequate  to describe both $\Phi(C)$ and $\vec{n}(C,t)$:
there is a fictitious spin 1/2 particle living on $C$, the polarization of which gives exactly
$\vec{n}(C,t)$; the wave function of the particle is defined for given gauge background
uniquely up to the phase and change of contour orientation is equivalent to time reversal operator
applied to the spinor; the particle evolution along $C$ is cyclic,
initial and final states differ only by phase and this phase is the magnitude of the flux
penetrating $C$. On the lattice the difference is that the flux direction
(wave function of spinning particle) is known only at lattice sites $x\in C$.
Moreover, the orientation of all elementary plaquettes is fixed to be the canonical one.

\subsection{Non-Abelian Stokes Theorem}
\label{NAST}
\noindent
The last ingredient which we need to complete the program outlined in sec.~\ref{program}
is the non-Abelian Stokes theorem derived recently by one of us~\cite{Gubarev:2003ij}.
Although the results of Ref.~\cite{Gubarev:2003ij} are applicable almost literally,
let us review them in order to introduce the notations and comment on the differences
with present work.

\begin{figure}
\centerline{\psfig{file=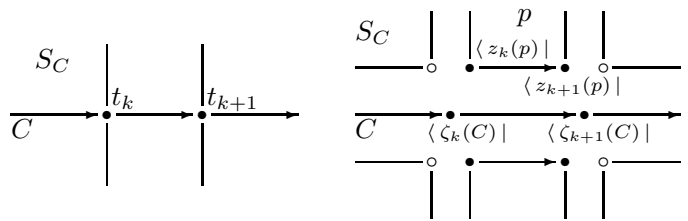,width=0.5\textwidth,silent=,clip=}}
\caption{Segment of the Wilson loop $W(C)$ in the original and ribbon-like representation.
The operator in between solid blobs is $U_{k,k+1}$, Eq.~(\ref{link}).}
\label{fig2}
\end{figure}

Therefore consider the Wilson loop $W(C)$, segment of which is shown by straight horizontal line
on Figure~\ref{fig2}, and the surface $S_C$ bounded by $C$, which is to the top of contour on the same figure.
According to what had been said above  we assign to every plaquette $p\in S_C$ and
Wilson loop itself the corresponding flux magnitudes $\Phi(p)$, $\Phi(C)$ and the instantaneous flux directions
$\bra{z_k(p)}$, $\bra{\zeta_k(C)}$ correspondingly (plaquette vertices are followed according to the
orientation induced by $C$ while the states $\bra{z_k(p)}$ are constructed in accordance with the canonical
orientation). It is convenient to use the graphical ribbon-like representation in which all plaquettes
and Wilson loop contour are slided apart (Figure~\ref{fig2}). Let us denote
\beq
\label{link}
U_{k, k+1} ~=~ \Pexp\,i\int^{t_{k+1}}_{t_k} \!\!\! A(\tau)\,d\tau
\eeq
and consider the matrix element
\beq
\label{matrix-element}
\bra{\zeta_k(C)}\, U_{k, k+1} \,\ket{\zeta_{k+1}(C)} ~=~ const \cdot e^{\,i\,\phi_{k,k+1}(C) }\,,
\eeq
where $const$ is some real positive number which is irrelevant. According to (\ref{flux2})
\beq
\label{total-flux}
\Phi(C) ~=~ \left[\,\sum_k \phi_{k,k+1}(C)\,\right] \,\mathrm{mod}\,2\pi\,.
\eeq
The important observation of Ref.~\cite{Gubarev:2003ij} is that the matrix element
(\ref{matrix-element}) might be calculated in $\bra{z_k(p)}$ basis provided that the
relative orientation of plaquette and Wilson loop fluxes is taken into account
\beqn
\bra{\zeta_k(C)} U_{k, k+1} \ket{\zeta_{k+1}(C)} = const \cdot \braket{\zeta_k(C)}{z_k(p)} \times \nonumber \\
\label{basis-change}
\times \, \bra{z_k(p)} U_{k, k+1} \ket{z_{k+1}(p)} \braket{z_{k+1}(p)}{\zeta_{k+1}(C)}\,.{~~~~}
\eeqn
The equality (\ref{basis-change}) was shown in Ref.~\cite{Gubarev:2003ij} in the matrix form.
Here we note that Eq.~(\ref{basis-change}) follows from its invariance under (\ref{rephasing})
and the unitarity of the evolution operator (\ref{link}). In fact, the relations similar to (\ref{basis-change})
are well known in quantum mechanics~\cite{pancharatnam}
(see, e.g.~\cite{Aharonov:1987gg,bhandari,Benedict} for details).
In particular, Refs.~\cite{bhandari,Benedict} showed the importance and physical significance
of the geodesic interpolation used in~\cite{Gubarev:2003ij}.

\begin{figure}
\centerline{\psfig{file=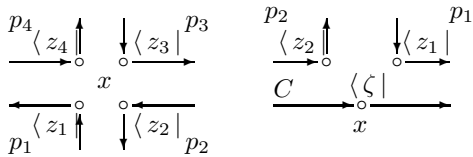,width=0.35\textwidth,silent=,clip=}}
\caption{$\Omega_x$ (left)  and $\gamma_x$ (right) terms in Eq.~(\ref{NAST-lattice}). Arrows correspond
to the orientation induced by $C$.}
\label{fig3}
\end{figure}

Applying Eqs.~(\ref{matrix-element}), (\ref{total-flux}), (\ref{basis-change}) repeatedly for
every link of $S_C$ one gets the non-Abelian Stokes theorem
\beqn
\Phi(C) = \sum\limits_{p\in S_C} I(p)\,\Phi(p)\,
        + \sum\limits_{x\in S_C} \Omega_x + \sum\limits_{x\in C} \gamma_x
        + \, 2\pi k(S_C)\,, \nonumber \\
\label{NAST-lattice}
k(S_C)\in Z\,, {~~~}
\eeqn
where $\Phi(p)$ is the plaquette flux, $1/2\tr W(p) = \cos\Phi(p)$, and the factors $I(p) = \pm 1$
are analogous to the usual incidence numbers in the differential geometry~\cite{Dubrovin}: $I(p) = 1$ if vertices
of the plaquette $p$ are followed in the canonical order and $I(p)=-1$ overwise. The remaining terms
are illustrated on Figure~\ref{fig3}. In particular, 
\beq
\label{omega-x}
\! \Omega_x \! = \mathrm{arg}[
\braket{z_1}{z_2} \braket{z_2}{z_3} \braket{z_3}{z_4} \braket{z_4}{z_1}]
\,\mathrm{mod}\, 2\pi
\eeq
is the oriented area of spherical quadrilateral polygon \footnote{
In our normalization  the total area of unit two-dimensional sphere is $2\pi$.
} (solid angle) in between the flux directions 
on the plaquettes $p_1,...,p_4$. It is known in quantum mechanics as Bargmann invariant~\cite{Bargmann:1964zj}
for the particle's wave functions (see, e.g.~\cite{Rabei:1999vu,Samuel:1997te} for review).
Physically $\Omega_x$ accounts for the difference of flux orientations on the plaquettes sharing
the same point $x$. The third term
\beq
\label{gamma-x}
\gamma_x ~=~ \mathrm{arg}[
\braket{\zeta}{z_1} \braket{z_1}{z_2} \braket{z_2}{\zeta}] \,\mathrm{mod}\, 2\pi
\eeq
equals to the oriented area of spherical triangle constructed from the Wilson loop flux
direction at $x$ and the flux orientations of two plaquettes $p_1,p_2\in S_C$ touching 
$C$ and sharing the point $x$. Eq.~(\ref{gamma-x}) is again the Bargmann invariant for the wave
functions of three particles living on $C$, $p_1$ and $p_2$. 

Note that we have omitted the mod $2\pi$ operation on the r.h.s. of Eq.~(\ref{NAST-lattice})
and wrote instead the additional $2\pi k(S_C)$ term, such that $\Phi(C)\in (0\,;\,\pi)$.
It is clear that $k(S_C)$ is not vanishing in general and is analogous to the Dirac string contribution
in the Abelian Stokes theorem applied for compact U(1) gauge fields~\cite{Polyakov:1975rs,DeGrand:1980eq}
(see \cite{Peskin:1977kp,Haymaker:1998cw} for review and further references).
This is in accordance with the discussion in sec.~\ref{violation-intro}, where we noted
that the SU(2) gauge model is intrinsically compact and is similar to compact photodynamics
in this respect. However, in the non-Abelian case the non-zero $k(S_C)$ could come from
either of three terms in Eq.~(\ref{NAST-lattice}). In particular, the Dirac string contribution
$k(S_C)\ne 0$ does not necessary corresponds to the singular elementary non-Abelian flux (singular
field strength). It could equally come from $\Omega_x$, $\gamma_x$ terms which are genuine non-Abelian
contributions.

Note that Eq.~(\ref{NAST-lattice}) is not only invariant under SU(2) gauge transformations, it also
remains intact with respect to local (gauge) rephasing (\ref{rephasing}) (this U(1) gauge symmetry
is crucial for the dual representation considered in Ref.~\cite{Chernodub:2000rg}).
We are in haste to add however, that this does not concern the $2\pi k(S_C)$ term. As might be expected
the Dirac string contribution is not invariant with respect to either of the symmetries.
Eq.~(\ref{NAST-lattice}) could be illustrated nicely in the particular case
of pure Abelian gauge background. In the Abelian limit all fluxes become aligned, but their directions
could be opposite. For anti-aligned flux directions the Bargmann invariants (\ref{omega-x}),(\ref{gamma-x})
become strictly speaking undefined. For instance, the area of the spherical triangle (\ref{gamma-x})
is undefined when two of its vertices are at the north pole of the two-dimensional sphere
while the third one is at the south pole. 
However, we could avoid this degenerate case by changing simultaneously the sign of both
$\vec{n}(p,t)$ and $\Phi(p)$ which does not affect the parametrization (\ref{wilson}).
The flux magnitude becomes not positively definite and the incidence coefficients could be
absorbed into the definition of $\Phi(p)$. Then the second and third terms, which
account for the flux rotation in color space, vanish  and Eq.~(\ref{NAST-lattice}) becomes
identical to the usual Abelian Stokes theorem.

To summarize, the flux $\Phi(C)$ could be represented almost entirely in terms of local physically
observable contributions coming from the arbitrary surface $S_C$ bounded by $C$.
The point of crucial importance is that all these terms are ``almost total differentials'':
without mod $2\pi$ operation both the plaquette flux (\ref{total-flux}) and the Bargmann
invariants (\ref{omega-x}), (\ref{gamma-x}) would become an exact 2-forms.
The adequate graphical language to account for all terms is the ribbon-like representation
in which all plaquettes and Wilson contour are slided apart.
The only troublesome contribution is the last one in Eq.~(\ref{NAST-lattice})
which explicitly depends upon the color orientation of the flux $\Phi(C)$ itself.
In the next section we analyze the arbitrariness of $S_C$ and $\gamma$-angles dependence
of Eq.~(\ref{NAST-lattice}).

\subsection{Non-Abelian Bianchi Identities}
\label{bianchi}
\noindent
To complete the program outlined in sec.~\ref{program} consider the surface independence
of the non-Abelian Stokes theorem (\ref{NAST-lattice}). As one could expect
the requirement of surface independence reduces to Eq.~(\ref{bianchi-NAST}).
On the other hand, the non-Abelian Stokes theorem~(\ref{NAST-lattice})
applied formally to closed surface $S_0$ gives
\beq
\label{bianchi-lattice}
\sum\limits_{p\in S_0} I(p)\,\Phi(p) ~+~ \sum\limits_{x\in S_0} \Omega_x ~=~ 2\pi\,q(S_0)\,,
\eeq
where the integer $q(S_0)$ is not vanishing in general and is discussed below.
Since Eq.~(\ref{bianchi-lattice}) is one of the central points of our work let us explicitly
rederive it starting from Eqs.~(\ref{kiskis}), (\ref{NAST-lattice}).

\begin{figure}
\centerline{\psfig{file=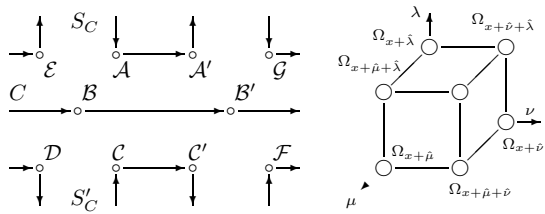,width=0.4\textwidth,silent=,clip=}}
\caption{Left: the non-Abelian Stokes theorem in application to the closed contour
$C$ which bounds two distinct surfaces $S_C$, $S_C'$, $\delta S_C=\delta S_C' = C$.
Arrows indicate the order of plaquette vertices induced by the orientation of $C$.
Right: the non-Abelian Bianchi identities for single lattice cube (see the text).}
\label{fig4}
\end{figure}

Consider Eq.~(\ref{kiskis}) for some closed contour $C$
\beq
\label{kiskis-double}
U( C \circ C^{-1}) = U(C) \, U(C^{-1}) = U(C) \, U^{-1}(C)  = 1\,,
\eeq
part of which is shown on Figure~\ref{fig4}. There are two distinct surfaces $S_C$, $S_C'$
shown to the top and bottom of the contour with orientations induced by $C$.
The non-Abelian Stokes theorem (\ref{NAST-lattice}) applied for $S_C$ and $S_C'$ leads to
\beq
\label{bianchi-S_C}
\! \Phi_{S_C} \! = \!\!  \sum\limits_{p\in S_C} \!\!  I(p) \Phi(p) + \!\!\! \sum\limits_{x\in S_C} \!\! \Omega_x
+ \! \sum\limits_{x\in C} \gamma_x(S_C) + 2\pi\,k(S_C)
\eeq
and analogous equation for $\Phi_{S_C'}$. The surface independence requires that
$\Phi_{S_C} = \Phi_{S_C'}$ and therefore
\beqn
\sum\limits_{p\in S_0} I(p) \Phi(p) +
\sum\limits_{\stackrel{x\in S_0}{x\notin C}} \Omega_x +
\sum\limits_{x\in C} [\gamma_x(S_C)-\gamma_x(S_C')] = \nonumber \\
\label{bianchi-1}
= 2\pi[ k(S_C') - k(S_C) ] = 2\pi\,q(S_0)\,.{~~~}
\eeqn
Here $S_0 = S_C \cup \tilde{S}_C'$ and $\tilde{S}_C'$ is just the $S_C'$ 
taken with reversed orientation due to which the terms
$\sum_{p\in S_C'}I(p)\Phi(p)$, $\sum_{x\in S_C'}\Omega_x$ changed sign in Eq.~(\ref{bianchi-1}).
Consider the $\gamma$-angles contribution in (\ref{bianchi-1}) coming from points
$\mathcal{B}, \mathcal{B}' \in C$
and let $\Delta(\mathcal{A}\mathcal{B}\mathcal{C})$ denotes the Bargmann invariant (\ref{gamma-x})
for spinor wave functions at the points $\mathcal{A}$, $\mathcal{B}$, $\mathcal{C}$.
In particular, $\gamma_{\mathcal{B}}(S_C) = \Delta(\mathcal{A}\mathcal{E}\mathcal{B})$
and similarly for other $\gamma$-angles.
We note that  one and the same unitary operator transforms $\mathcal{A}\to\mathcal{A}'$,
$\mathcal{B}\to\mathcal{B}'$, $\mathcal{C}\to\mathcal{C}'$. In other words the color directions
of the fluxes at these points are rotated by one and the same rotation matrix.
However, the Bargmann invariant being the area of the spherical triangle is unchanged when sphere
is rotated. Therefore, the following identity holds
\beq
\label{bargmann}
\Delta(\mathcal{A}\mathcal{B}\mathcal{C}) ~-~ \Delta(\mathcal{A}'\mathcal{B}'\mathcal{C}') ~=~ 0\,.
\eeq
It is clear that when Eq.~(\ref{bargmann}) taken for each link of $C$ is added 
to the l.h.s. of (\ref{bianchi-1}) the total $\gamma$-angles contribution becomes
\beq
\sum\limits_{x\in C} [\gamma_x(S_C) - \gamma_x(S_C')] ~=~ \sum\limits_{x\in C} \Omega_x\,,
\eeq
where the orientation change of $S_C'$ in the inclusion $S_0=S_C \cup \tilde{S}_C'$ is crucial.
For instance, $\Omega(\mathcal{B})$ is given by $\Delta(\mathcal{A}\mathcal{E}\mathcal{D}\mathcal{C})$
and does not depend at all on contour $C$. We conclude therefore that Eq.~(\ref{bianchi-lattice})
is the consistency requirement for the non-Abelian Stokes theorem (\ref{NAST-lattice})
to be independent on the surface. But the point is that Eq.~(\ref{bianchi-lattice})
is more than the consistency condition. As we have argued in sec.~\ref{program},
Eq.~(\ref{bianchi-lattice}) being applied to the infinitesimal cube is in fact
the lattice implementation of the non-Abelian Bianchi identities and is illustrated on
Figure~\ref{fig4} (right). It is clear that the integer $q(S_0)$
is the magnetic charge discussed in sec.~\ref{program}.
Therefore, the non-Abelian Stokes theorem (\ref{NAST-lattice})
which refers explicitly to the physically observable field strength allows to formulate
the non-Abelian Bianchi identities on the lattice and to study their violation
in gauge invariant terms.

\subsection{Discussions}
\label{discussions}
\noindent
This section is devoted to general notes concerning the Bianchi identities
and the magnetic charge definition. We do not pretend on the exhaustive
treatment, of course. However the following items seem to be worth mentioned:

{\it i)} The SU(2) gauge invariance of the magnetic charge
is evident from the fact that each term on the l.h.s. of (\ref{bianchi-lattice})
is SU(2) gauge invariant by construction. The U(1) gauge invariance (\ref{rephasing})
of Eq.~(\ref{bianchi-lattice}) is also obvious. One could argue
that this Abelian symmetry is artificial and is only due to our intent to represent
the non-Abelian flux direction in terms of the fictitious spinning particle. However, we do think
that the U(1) invariance of (\ref{bianchi-lattice}) might be relevant. Indeed, the interpretation of the Wilson
loop defining equation (\ref{evolution}) in quantum mechanical language is natural
and forces us to concentrate on the phase differences of wave functions (see, e.g.,
Eqs.~(\ref{eigen}), (\ref{omega-x}), (\ref{gamma-x})), not on their concrete phases.
Moreover, it allows to use the machinery related to the line
bundle structure of quantum mechanics, mathematical foundations of geometrical phases 
and Bargmann invariants. In this respect the U(1) symmetry appears naturally and
is inherent to our approach (it had been also discussed
although in different context in Refs.~\cite{Chernodub:2000wk,Chernodub:2000rg}).

{\it ii)}  What was also crucial for our construction is the canonical orientation
of elementary lattice plaquettes. We discussed this in details in sec.~\ref{chromomagnetic-fields}
and concluded that in order to deduce the field strength from the infinitesimal Wilson loops
some canonical ordering must be introduced. It is true that in most cases the concrete ordering
prescription does not matter since the usually considered quantities do not depend on it.
For instance, the gauge action is insensitive to plaquette orientations, but this is certainly because
the action is even in the field strength. As far as the magnetic fields
are concerned their unambiguous definition is only possible with some
canonical ordering prescription, overwise the components of $F_{\mu\nu}$
could be determined only up to the sign even in the Abelian theory.
However, it is clear that the ordering is
not unique and although there are only few possibilities to choose from,
the dependence of Eq.~(\ref{bianchi-lattice}) on the particular choice should be
investigated separately. In this work we stuck with the conventional canonical ordering
described above, the ordering dependence will be investigated elsewhere.

{\it iii)} As we have noted already it is natural to describe the non-Abelian Stokes theorem
(\ref{NAST-lattice}) in the ribbon-like graphical representation in which the theorem
becomes essentially Abelian-like. In other words the non-Abelian nature of the theory
is traded for the complicated geometry. Therefore the ribbon-like representation
is actually not only the convenience. Once we could unambiguously assign each term in
Eqs.~(\ref{NAST-lattice}), (\ref{bianchi-lattice}) to a particular geometrical object
it is natural to ask whether these objects form a self-contained cell complex.
For the non-Abelian Stokes theorem the answer is ``no'' because each Wilson contour requires
the introduction of its own set of triangles
(e.g., $\mathcal{A}\mathcal{B}\mathcal{E}$ on Figure~\ref{fig4})
to which the $\gamma$-angles are to be ascribed.  But the non-Abelian Bianchi identities
do indeed allow the introduction of specific cell complex in which every term on
the l.h.s. of Eq.~(\ref{bianchi-lattice}) is unambiguously assigned to
the particular 2-dimensional cell. Moreover, Eq.~(\ref{bianchi-lattice}) could then
be interpreted as usual coboundary operator acting on 2-cochains.
Note that the above reasoning resemble slightly the dual gravity-like representation
of SU(2) gluodynamics~\cite{Lunev,Ganor:1995em,Bauer:1994hj,Haagensen:1994sy,Diakonov:2001xg}.
We stress that this approach is not only the mathematical convenience. 
In fact it is the only way to analyze the structure of Eq.~(\ref{bianchi-lattice})
at finite lattice spacing. In particular, it allows to show that the magnetic
charge is closely related to the degenerate points (\ref{degenerate})
mentioned in sec.~\ref{degeneracy} (this is the topic of the next section).
Here we note that the cell complex underlying Eq.~(\ref{bianchi-lattice}) is
described in Appendix the results of which are used in the next section.

{\it iv)} It seems to be instructive to start from Eq.~(\ref{bianchi-lattice}),
expand it in powers of the lattice spacing and get the Bianchi identities (\ref{bianchi-D4}), (\ref{bianchi-D3})
in the continuum limit. However, we failed to implement this program. 
As far as we can see the reason is two-fold. First, the original problem (\ref{magnetic-charge})
was posed quite differently from what could be expected in the continuum.
Indeed, our primary goal was to determine the magnetic charge and we intentionally refused
to consider its gauge dependent color orientation.
The manifestation of this could be seen by comparing Eqs.~(\ref{bianchi-D4}), (\ref{bianchi-D3}) with (\ref{bianchi-lattice}):
while the former is in the adjoint representation and is vector in the color space the later
is gauge invariant and is just one equation. Therefore it is {\it a priori} unclear 
how one could get (\ref{bianchi-D4}), (\ref{bianchi-D3}) from (\ref{bianchi-lattice}) even in the limit of vanishing
lattice spacing. On the other hand, Eq.~(\ref{bianchi-lattice}) follows rigorously
from (\ref{kiskis}) and we have no doubt that Eq.~(\ref{bianchi-lattice}) indeed expresses the Bianchi identities
on the lattice. Secondly, as we argue in the item below (see also the next section) the discussion
of Eq.~(\ref{bianchi-lattice}) in the continuum limit is indispensable from the consideration
of the degenerate points (\ref{degenerate}).

{\it v)} Let us qualitatively consider what happens with the magnetic charge (\ref{bianchi-lattice})
in the extreme weak coupling limit. The plaquette fluxes do not play any role since
they are highly suppressed by the action. Therefore Eq.~(\ref{bianchi-lattice}) simplifies
\beq
\label{bianchi-lattice-weak}
\sum\limits_{x\in \delta c} \Omega_x ~=~ 2\pi\,q(c)\,,
\eeq
where $c$ is elementary lattice cube. Note that the magnetic charge in not directly suppressed
by the action and therefore there seems to be no reasons for it to die out in the continuum limit.
Moreover, it is clear from (\ref{bianchi-lattice-weak}) that
the non-zero $q(c)$ is due to the particular distribution of the chromomagnetic field directions
and is almost insensitive to the magnitude of the elementary fluxes. Indeed, each
$\Omega_x$ depends only on the flux directions and not on their magnitudes. In the next section we show that
the non-zero r.h.s. of Eq.~(\ref{bianchi-lattice-weak}) in the continuum limit
indicates that at this point the chromomagnetic fields are degenerated
and the particular determinant constructed from $E^a_i$, $B^a_i$ vanishes.

\section{Chromomagnetic Fields Degeneracy}
\noindent
In this section we analyze the points of chromomagnetic fields degeneracy introduced
in sec.~\ref{degeneracy}. First we review the essential facts known in the continuum
and then turn to the lattice definitions.

\subsection{Preliminaries}
\noindent
In four dimensions the points of degeneracy of the chromomagnetic fields are defined by
\beq
\label{deg-null}
\mathrm{det}\, T ~=~ 0\,,
\eeq
\beq
\label{deg-operator}
T^{ab}_{\mu\nu} ~=~ \varepsilon^{abc}\, \tilde{F}^c_{\mu\nu} ~=~
\frac{1}{2} \,\varepsilon^{abc}\, \varepsilon_{\mu\nu\lambda\rho} F^c_{\lambda\rho}\,.
\eeq
As we noted in sec.~\ref{degeneracy} the physical significance of the points (\ref{deg-null})
crucially depends on the dimensionality. Indeed, in D=3 the operator coupled to the gauge
potentials $A^a_k$ in the Bianchi identities (\ref{bianchi-D3}) is
\beq
\label{deg-operator-D3}
{T^{ab}_k}_{(3D)} ~=~ \varepsilon^{abc}\, B^c_k ~=~ \frac{1}{2} \,\varepsilon^{abc}\,\varepsilon_{kij}\, F^c_{ij}
\eeq
and, in fact, is $3 \times 9$ matrix for which the determinant is undefined.
We could at best consider the rank of the matrix (\ref{deg-operator-D3}) and clearly
\beq
\mathrm{rank}\, T_{(3D)} ~ < ~ 3\,,
\eeq
since $B^a_k$ is always the eigenvector with zero eigenvalue. We conclude therefore
that in D=3 the very notion of chromomagnetic fields degeneracy is uncertain.

In four dimensions the $\mathrm{det}\,T$ was calculated long
ago~\cite{Roskies:1976ux,Halpern:1977fw,Deser:1976iy}:
\beq
\mathrm{det}\, T ~ \propto ~ \mathrm{det}\, K\,,
\eeq
\beq
\label{deg-K}
K_{\mu\nu} = K_{\nu\mu} =
\frac{1}{3}\,\varepsilon^{abc} \,\tilde{F}^a_{\mu\rho}\, F^b_{\rho\lambda}\, \tilde{F}^c_{\lambda\nu}\,,
\eeq
\beqn
\! K_{00} = 2 \,\mathrm{det}\,B\,,{~} &
K_{ik} = \frac{1}{2} B^a_{\{i} \, Q^a_{k\}}\,,{~} &
K_{0i} = B^a_i \,\varepsilon^{abc}\, E^b_k \,B^c_k\,, \nonumber \\ \nonumber \\
E^a_i = F^a_{0i}\,,{~} &
B^a_i = \frac{1}{2}\,\varepsilon_{ijk}\,F^a_{jk}\,,{~} &
Q^a_k = \varepsilon^{abc}\,\varepsilon_{kij}\, E^b_i\,E^c_j\,, \nonumber
\eeqn
where curly braces denote symmetrization $B_{\{i} Q_{k\}} = B_i Q_k + B_k Q_i$.
It is important that each element of $K_{\mu\nu}$ is
gauge invariant determinant constructed in terms of $E^a_i$ and $B^a_i$.
In particular, the off-diagonal elements are of the form 
$\mathrm{det}[E_i E_k B_k]$, $\mathrm{det}[B_i E_k B_k]$, where no summation in $k$
is implied and $\mathrm{det}[e_1 e_2 e_3]$ is understood as the determinant of the column matrix
constructed from color vectors $\vec{e}_1,\vec{e}_2,\vec{e}_3$. Note that the off-diagonal
elements vanish identically in (anti)self-dual sectors~\cite{Halpern:1977ia}. However, our aim is not to
analyze Eqs.~(\ref{deg-null})-(\ref{deg-K})
in their generality. Rather we would like to show that the lattice Bianchi identities
naturally lead to the same determinants (\ref{deg-K}).
In particular, in the next section we show that the magnetic charge (\ref{bianchi-lattice})
is ultimately related to the zeros of these determinants and hence to the degenerate points (\ref{deg-null}).

\subsection{${\bf \mathrm{\bf det}\,B=0}$ on the Lattice}
\label{detB}
\noindent
In this section we consider first the three dimensional case which is much simpler geometrically.
The results remain valid in four dimensions, but in D=4 there are important differences as well.

It was noted in sec.~\ref{discussions} that the only reliable and rigorous way
to analyze Eq.~(\ref{bianchi-lattice}) at finite lattice spacing is to consider
the specially crafted cell complex for which Eq.~(\ref{bianchi-lattice}) is
the coboundary operation. The existence and structure of this cell complex 
could be inferred by noting that the non-Abelian Stokes theorem and the Bianchi identities
on the lattice are described most naturally in the ribbon-like graphical representation.
Starting from this the cell complex could be completely constructed
(see Appendix). The advantage of this approach is that
it is rather formal. Once we were able to assign a gauge invariant numbers
(magnitudes of elementary fluxes and their relative orientations expressed in terms
of Bargmann invariants) to each 2-dimensional cell, all we have to do is to consider the coboundary
operator $d: \mathbb{C}^2\to\mathbb{C}^3$, where $\mathbb{C}^k$ is the $k$-skeleton.
For every lattice cube the action of $d$ is
equivalent to the Bianchi identities (\ref{bianchi-lattice}) by construction and
hence $d$ assigns the corresponding magnetic charge to each lattice cube.
However, the additional $\Omega$-angles contribution implies that
the geometry of the cell complex is not (hyper)cubical.
In particular, the 3-skeleton $\mathbb{C}^3$ is larger than the union of lattice 3-cubes.

\begin{figure}
\centerline{\psfig{file=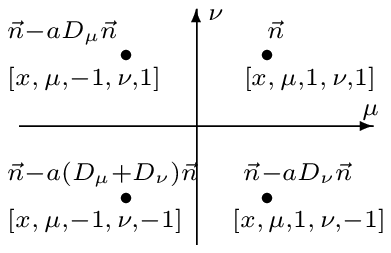,width=0.25\textwidth,silent=,clip=}
\hspace{0.02\textwidth}
\psfig{file=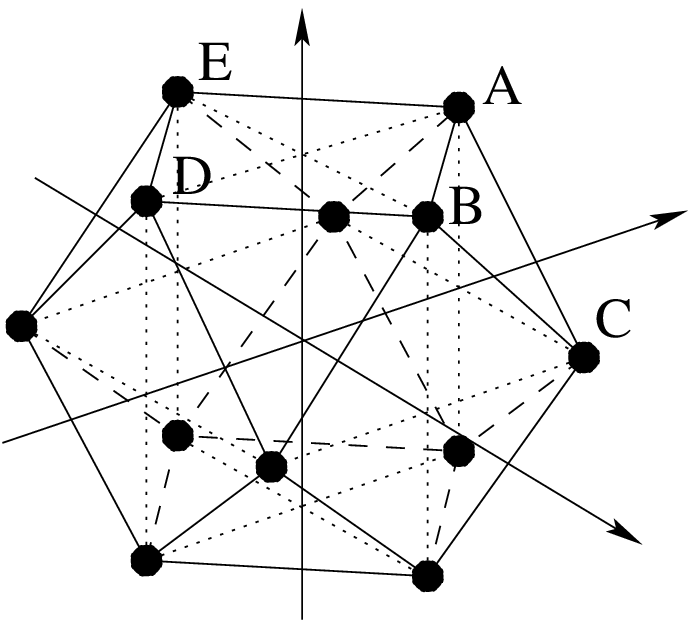,width=0.2\textwidth,silent=}}
\caption{Left: the flux directions in the plane $(\mu,\nu)$ around point $x$ in the
weak coupling limit, eq~(\ref{weak-couling}). Right: the only non trivial 3-cell in D=3 (see the text).}
\label{fig5}
\end{figure}

On the other hand, there is formally no difference between different 3-cells of the complex.
In particular, one can show that $d: \mathbb{C}^2\to\mathbb{C}^3$ always assigns
an integer number to every 3-cell. It is true that some of this ``new'' 3-cells are trivial
and the corresponding magnetic charge is always zero.
However, there exist the non-trivial cases as well (see Appendix) one of which
(and the only one in D=3) is illustrated on Figure~\ref{fig5} (right).

Consider some point $x$ on the original D=3 lattice together with 12 plaquettes and 8 cubes which
share this point. Eq.~(\ref{bianchi-lattice}) applied to each cube
forces us to take into account 8 triangles at cube's corners (cf. Figure~\ref{fig4})
and to assign the corresponding Bargmann invariants $\Omega_i$, $i=1,...,8$ to each triangle. Figure~\ref{fig5} (right)
shows the triangles around point $x$ coming from different cubes. Note that all 8 triangles
are properly oriented. By the same token  one concludes that 6 squares, e.g. $ABDE$, are also valid 2-cells
of the cell complex and are equipped with the corresponding
Bargmann invariants $\Delta_i(x)$, $i=1,...,6$. Then it is clear that
the application of $d: \mathbb{C}^2\to\mathbb{C}^3$
to the set of 2-cells on Figure~\ref{fig5} assigns a well defined and gauge invariant
integer number to the 3-cell shown on that figure:
\beq
\label{site-charge}
2\pi\,\tilde{q}(x) ~=~ \sum\limits_{i=1}^8 \Omega_i(x) ~+~ \sum\limits_{i=1}^6 \Delta_i(x)\,.
\eeq
Formally  it is just the same magnetic charge we have considered so far, but now it is ascribed to the site
of the original lattice. We are confident that the magnetic charges
in the lattice cubes correspond to the Bianchi identities violation.
But what is violated in the lattice sites?

To answer this question we expand (\ref{site-charge}) in powers of the lattice spacing.
However, it is worth to mention that this expansion is not the usual one.
In particular, it would be plainly wrong to look for $O(a^3)$ terms since
the integer number on the l.h.s. of Eq.~(\ref{site-charge})
does not depend at all on the lattice spacing.
Therefore in the weak coupling expansion we should look for $a$-independent
contributions or, better to say, to look for the conditions for $a$-independent terms to appear.

In fact, all necessary relations were derived in Ref.~\cite{Gubarev:2003ij}.
In particular, consider four plaquettes in the same plane which share the point $x$,
Figure~\ref{fig5} (left). To the leading order the color directions of the fluxes at point $x$ are given by
\beq
\label{weak-couling}
\begin{array}{ccl}
x + \hat{\mu} + \hat{\nu} & : & \vec{n} + O(a^2)\,, \\
x + \hat{\mu} - \hat{\nu} & : & \vec{n} - a\,D_{\nu}\vec{n} + O(a^2)\,, \\
x - \hat{\mu} + \hat{\nu} & : & \vec{n} - a\,D_{\mu}\vec{n} + O(a^2)\,, \\
x - \hat{\mu} - \hat{\nu} & : & \vec{n} - a\,(D_{\mu} + D_{\nu})\vec{n} + O(a^2)\,,
\end{array}
\eeq
where we have denoted $\vec{n}=\vec{n}_{(\mu\nu)}(x)$ for brevity. We conclude therefore
that in the weak coupling limit the three points $A,B,C$ (Figure~\ref{fig5}, right)
are distinguished: the flux directions assigned to them are in general independent
and coincide with color direction of the particular component of $\vec{F}_{\mu\nu}$.
The flux directions in all other vertices are obtainable by infinitesimal variation
of the flux direction in one of the points $A,B,C$.

Recall now that the Bargmann invariant assigned to each triangle and square
is the oriented solid angle between the corresponding flux directions.
It follows then that the contribution of all squares is always of order $O(a^3)$
and is negligible. As far as the triangles are concerned they also give terms
of order $O(a^3)$ unless the fluxes at points $A,B,C$ 
become linearly dependent. In this case the corresponding Bargmann invariant
could be $\pm\pi + O(a)$ and the order $O(a)$ variation of the fluxes
at various vertices is enough to change it by $2\pi$.
It is clear that only in this degenerate case the non-zero l.h.s. of
(\ref{site-charge}) is at all possible.
On the other hand, the flux directions at the points $A,B,C$ in the weak coupling limit
are given by the corresponding chromomagnetic field components $\vec{B}_k$. We conclude
therefore that the non-vanishing magnetic charge (\ref{site-charge})
implies that the chromomagnetic fields are degenerate at this point
\beq
\label{q-detB}
\tilde{q}(x) \ne 0 \,\,\Rightarrow \,\,\, \mathrm{det}\,B(x) = 0\,.
\eeq
Note that the statement could not be reversed. For instance, in the case 
$\vec{B}_1 = \vec{B}_2 = \vec{B}_3$ both $\mathrm{det}\,B$ and $\tilde{q}$ vanish.

Eq.~(\ref{q-detB}) remains valid in four dimensions as well. The only distinction is that
now we have 4 different magnetic charges $\tilde{q}_\mu(x)$ labeled by the direction $\hat{\mu}$
dual to given 3-dimensional slice. In particular, the non-zero $\tilde{q}_\mu(x)$ implies that
one of the determinants $\mathrm{det}[B_1 B_2 B_3]$, $\mathrm{det}[B_1 E_2 E_3]$,
$\mathrm{det}[E_1 B_2  E_3]$, $\mathrm{det}[E_1 E_2 B_3]$ vanishes.
Note that these determinants are the diagonal entries of $K_{\mu\nu}$, Eq.~(\ref{deg-K})
and therefore
\beq
\label{q-K}
\tilde{q}_\mu(x) \ne 0 \,\,\Rightarrow \,\, 
K_{\mu\mu} = 0 \quad\mbox{(no sum over $\mu$)}\,.
\eeq
By symmetry considerations one expects that there should exist 3-cells for which the magnetic charge
indicates the zeros of 
$\mathrm{det}[E_1 E_2 E_3]$, $\mathrm{det}[E_1 B_2 B_3]$,
$\mathrm{det}[B_1 E_2 B_3]$, $\mathrm{det}[B_1 B_2 B_3]$.
It turns out that these cells are $\mathscr{D}^{(1)}(x,\mu,d_\mu)$ (see Appendix).
Indeed, the structure of $\mathscr{D}^{(1)}$ cells is such that the argumentation leading to (\ref{q-detB})
applies literally. Then the inspection of the flux directions assigned to vertices of $\mathscr{D}^{(1)}(x,\mu,d_\mu)$
shows that the non-zero magnetic charge of one of these 3-cells is the sufficient
condition for the particular determinant above to vanish.

As far as the off-diagonal elements of $K_{\mu\nu}$ are concerned, they are highly sensitive
to the  topological properties of the gauge fields. For instance,
$K_{12} = \mathrm{det}[B_2 E_2 E_3] - \mathrm{det}[B_1 E_1 E_3]$
vanishes in the (anti)self-dual sectors. It is possible to identify the 3-cells
which are related to the off-diagonal entries of $K_{\mu\nu}$ matrix.
Indeed, consider the diamond-like 3-cells $\mathscr{D}^{(2)}$ (see Appendix).
In the weak coupling limit the flux directions assigned to 4 plaquette corners
become essentially the same and coincide with the corresponding component of $\vec{F}_{\mu\nu}$.
Then the flux orientations ascribed to 3 pairs of opposite vertices of $\mathscr{D}^{(2)}$
are given by $\vec{E}_k$, $\vec{B}_k$, $k=1,2,3$.
Geometrically it is clear that for $\vec{E}_k = \pm \vec{B}_k$
the 3-cells $\mathscr{D}^{(2)}$ are highly degenerated and there is a good chance for
the coboundary operator $d: \mathbb{C}^2\to\mathscr{D}^{(2)}$ to give a non-zero magnetic charge.
However, we are still lacking the rigorous argumentation here.
One could only say (see also sec.~\ref{gamma_s}) that the $\mathscr{D}^{(2)}$ cells are indeed closely
connected to the topological properties of the gauge background. The relation of the present
approach to the gauge fields topology goes beyond the scope of the present publication and
will be investigated elsewhere.

To summarize, the non-Abelian Bianchi identities (\ref{bianchi-lattice})
could be interpreted as the coboundary operator $d: \mathbb{C}^2\to\mathbb{C}^3$
for the specific cell complex, the complicated geometry of which is the direct
consequence of the non-Abelian nature of the theory. Moreover, the operator $d$ considered
in its generality necessitates the consideration of gauge invariant magnetic charges
associated with various 3-cells. While the non-vanishing magnetic charge
in 3-dimensional cube implies the violation of the Bianchi identities, in other 3-cells
it is the sufficient condition for the particular determinant constructed from $E^a_i$,$B^a_i$ to vanish.
At finite lattice spacing these two types of magnetic charges are almost independent and should
be considered as such especially since they are geometrically distinct: the former are ascribed
to the lattice cubes, the later are assigned to the sites of the original lattice.
However, at vanishing lattice spacing the two types of magnetic charges become
closely interrelated (cf. Eq.~(\ref{bianchi-lattice-weak})): once the flux magnitude on the
elementary plaquettes becomes negligible everywhere the non-Abelian Bianchi identities could only be
violated at the degenerate points (\ref{q-K}).

\section{Numerical Experiments}
\label{numerics}
\noindent
It is true that the relevance of the above construction for the dynamics of the Yang-Mills fields
is not evident from the preceding presentation. However,
we specifically kept in mind  from very beginning the possibility to apply our approach in real lattice experiments.
In this section we describe the results of our numerical simulations.
The problem to be considered is whether the violation of the Bianchi identities and
the degeneracy of the chromomagnetic fields are physically significant.

The general setup is as follows. We simulate the SU(2) lattice gluodynamics in three
and four dimensions on the symmetric lattices with periodic boundary conditions.
The action we adopt initially (see below) is the standard Wilson action.
Until the sec.~\ref{gamma_c} the lattices we used are $16^3$ and $10^4$ with corresponding
$\beta$-ranges $[5.0;9.0]$ and $[2.2;2.8]$. Note that these parameters are partially
unphysical.  The purpose is to consider the behavior of the magnetic charges
(\ref{bianchi-lattice}), (\ref{q-detB}), (\ref{q-K})
in various circumstances, in particular, across the finite-volume phase transition.

The simplest and instructive quantities to study are the densities $\rho(\beta)$, $\tilde{\rho}(\beta)$
of the magnetic charges (\ref{bianchi-lattice}), (\ref{q-K}).
The density $\rho(\beta)$ is defined irrespectively of the space-time dimensionality
\beq
\label{rho-cube}
\rho(\beta) ~=~ \frac{1}{N_c}\,\sum\limits_c |q(c)|\,,
\eeq
where summation is over all lattice 3-cubes and $N_c$ is their total number. Evidently
$\rho(\beta)$ measures the fraction of points at which the non-Abelian Bianchi identities
are violated. The definition of $\tilde{\rho}(\beta)$ differs in D=3 and D=4. 
In three dimensions we have
\beq
\label{rho-site}
\tilde{\rho}(\beta) ~=~ \frac{1}{N_s}\,\sum\limits_s |\tilde{q}(s)|\,,
\eeq
where $s$ is the lattice site, $N_s$ is the total lattice volume and $\tilde{q}(s)$
was defined in sec.~\ref{detB}.
In D=4 there are several types of the magnetic charges $\tilde{q}$ and therefore
the definition (\ref{rho-site}) is ambiguous.
We take the symmetric definition which looks similar to (\ref{rho-site}):
$s$ denotes the 3-cell which is not the lattice cube and $N_s$ is the total number of these
cells. Physically $\tilde{\rho}(\beta)$ is the fraction of the lattice volume
occupied by zeros of various determinants, e.g. (\ref{q-detB}), (\ref{q-K}).

\begin{figure}
\centerline{\psfig{file=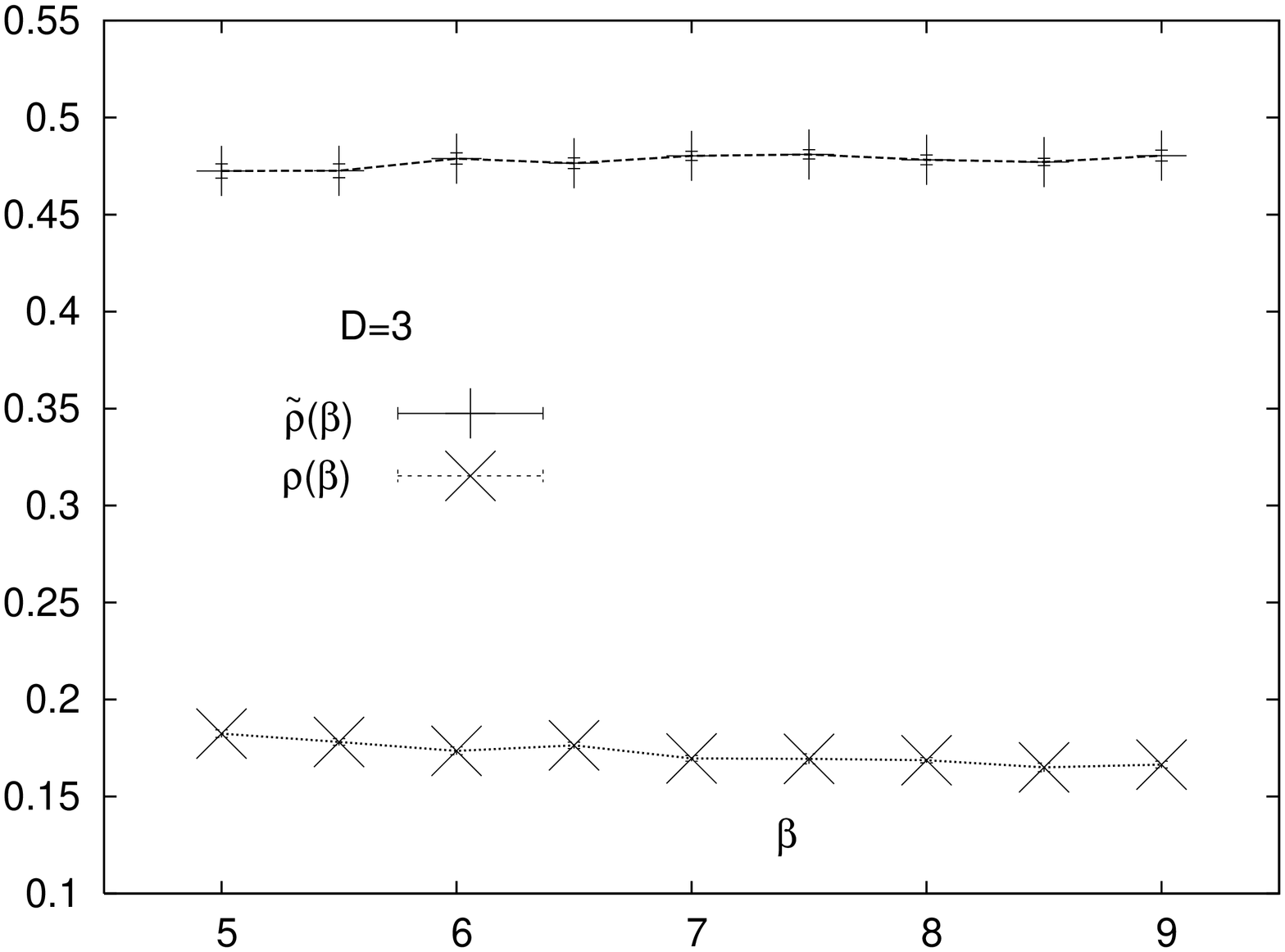,width=0.4\textwidth,silent=,angle=-90}}
\centerline{\psfig{file=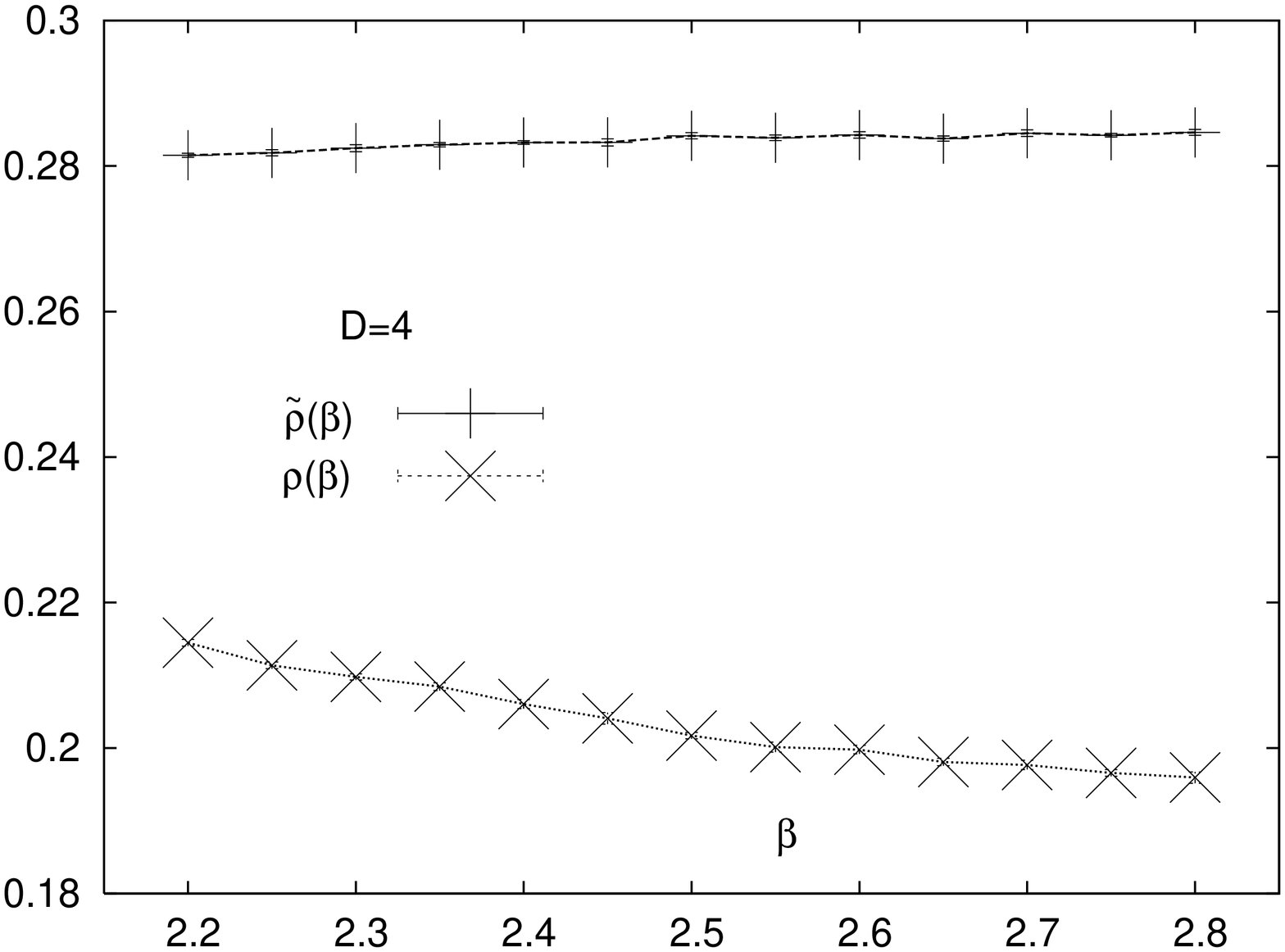,width=0.4\textwidth,silent=,angle=-90}}
\caption{The densities (\ref{rho-cube}), (\ref{rho-site}) versus $\beta$ coupling.
The lines are drawn to guide the eye.}
\label{fig6}
\end{figure}

The dependence of $\rho$ and $\tilde{\rho}$ on $\beta$ is shown on Figure~\ref{fig6}.
One can see that both densities are numerically similar in three and four dimensions and are
almost $\beta$-independent in accordance with general arguments of sec.~\ref{discussions}.
Indeed, the $\beta$-independence
of $\tilde{\rho}$ is certainly expected since there is no symmetry which could keep
the sign of the determinants (\ref{q-K}) fixed. In particular, the perturbation
theory gives the dominant contribution to the density $\tilde{\rho}(\beta)$.
On the other hand the $\beta$-independence of $\rho$ follows from the fact that
the violation of Bianchi identities is closely related to the zeros of the above determinants.
Therefore we come to the paradoxical conclusion
that the perturbation theory also saturates the density $\rho(\beta)$.

To resolve the problem we note that in the continuum limit
the Bianchi identities are formulated for elementary 3-volumes while
the determinants are defined at any particular point.
The corresponding construction on the lattice is essentially the same: the Bianchi identities
and the magnetic charge $q$ are ascribed to the elementary 3-cubes while the degeneracy
points and the charge $\tilde{q}$ are assigned to the lattice sites.
It is important that these charges are geometrically distinct on the lattice:
at arbitrary small but non-zero spacing there is $O(a)$
distance between them and they are defined on different 3-cells.
It turns out that on the lattice the magnetic charge at 3-cube
and anti-charge at the neighboring site may coexist with almost no additional 
action penalty (cf. Eqs.~(\ref{bianchi-lattice-weak}), (\ref{site-charge})).
Moreover, one can show that there could be no mechanism to prevent the creation
of these ultraviolet (UV) $q$-$\tilde{q}$ pairs since it would violate the gauge invariance.
Indeed, although the relative orientation of the fluxes is formally gauge invariant,
any restriction of it will effectively squeeze the non-Abelian fluxes into one
particular color direction. Then  it would be hardly possible to call the resulting
theory non-Abelian \footnote{
However, the restriction on the relative flux orientations seems to be an interesting
possibility to be investigated elsewhere.
}. Note that the UV pairs above are irrelevant from the continuum viewpoint. 
Indeed, there is no trace whatsoever of the ultraviolet $q$-$\tilde{q}$ pairs on the blocked
lattice with lattice spacing $N \cdot a$. 
At the same time the densities $\rho(\beta)$, $\tilde{\rho}(\beta)$ account for all the charges
$q$, $\tilde{q}$ on equal footing and therefore are dominated by the UV fluctuations. 

We conclude therefore that the densities $\rho(\beta)$ and $\tilde{\rho}(\beta)$ 
are not the appropriate observables on the unblocked lattices. 
They are dominated by the ultraviolet noise which is only due to the mismatch in the domain of
definition of the Bianchi identities and the degenerate points.
It seems that the only way to make sense of the densities $\rho$, $\tilde{\rho}$
is to consider them on the blocked configurations for which the ultraviolet noise
is gradually removed. However, our approach to the problem is different and is described below.

\subsection{Modification of the Action}
\noindent
As follows from the above presentation, the dynamics of $q$ and $\tilde{q}$ magnetic charges
is highly UV sensitive and the dominant configurations are small (at the scale of UV cutoff)
$q$-$\tilde{q}$ pairs. It seems that this observation forbids the discussion
of the significance of the Bianchi identities violation and the points of degeneracy
since it is impossible to separate the UV noise from physically relevant excitations.
Essentially the same problem exists in usual field theories, where the vacuum condensates
are commonly used to parametrize the non-perturbative effects.
The well known example is the gluon condensate $\langle\alpha_s (F^a_{\mu\nu})^2\rangle$
which is perturbatively divergent but its non-perturbative part is non-vanishing and is known to be
of major phenomenological importance.
The subtraction of the perturbative tail of various condensates is 
challenging and the usual approach is to subtract it order by order in the coupling constant.
However, we don't see any tractable way to do this in our case.

On the other hand, it is possible to reformulate slightly  the original problem. Instead
of trying to isolate the effects due to the UV $q$-$\tilde{q}$ pairs
we could equally ask what happens when the magnetic charges are partially
removed from the vacuum. Indeed, the definition of $q$ and $\tilde{q}$ charges is
local and gauge invariant. Therefore, nothing prevents us from modifying the Wilson action
to include the additional terms which could influence the dynamics of $q$, $\tilde{q}$ charges.
Since it is hardly possible to invent the additional well defined terms which are sensitive
to the UV dynamics only, we will study the following simplest modification
\beq
\label{action}
S = -\beta\sum\limits_p \Tr{2}U_p + \gamma \sum\limits_c |q(c)|
+ \tilde{\gamma} \sum\limits_s |\tilde{q}(s)|\,,
\eeq
where the first term is the standard Wilson action and $c$ denotes the elementary lattice cubes.
The last term in Eq.~(\ref{action}) has different interpretation in three and four dimensions.
In D=3 $s$ denotes the lattice sites and $\tilde{q}(s)$ is given by Eq.~(\ref{site-charge}).
In four dimensions the last term {\it a priori} depends on the concrete definition of the magnetic
charges $\tilde{q}$. As in the previous section we take the symmetric definition:
$s$ denotes the 3-cells which are not the lattice cubes and $\tilde{q}(s)$ is the corresponding
magnetic charge. It turns out that our results are almost insensitive to the particular
choice of the last term in Eq.~(\ref{action}), see sec.~\ref{gamma_s}.

The modified action is local and SU(2) gauge invariant. Indeed, from the defining
equations (\ref{bianchi-lattice}), (\ref{site-charge}) one can see that (\ref{action}) intertwines
the links which are at most two lattice spacings apart, while the gauge invariance follows by construction.
Then the universality suggests that the continuum limit of the model defined by (\ref{action})
should be the same as one for the model with the conventional Wilson action
(see also sec.~\ref{discussions2} for discussions).
On the other hand, the additional coupling constants $\gamma$, $\tilde{\gamma}$ allow to study
the effects which are due to the Bianchi identities violations and the degeneracy points.
The particular limit $\gamma \to \infty$ is of special interest since it corresponds to
the theory with nowhere violated Bianchi identities. As far as the $\tilde{\gamma}$ coupling
is concerned we are not so confident that the limit $\tilde{\gamma} \to \infty$
corresponds to a sensible theory. For instance, in D=3 the nowhere vanishing
$\mathrm{det}\,B = \mathrm{det}[B_1 B_2 B_3]$ implies that it is of the same sign everywhere,
which contradicts the perturbative expectations \footnote{
We're thankful to prof.~V.~I.~Zakharov for pointing this out.
} and probably violates $CP$ symmetry. At the same time the point $\gamma = \tilde{\gamma} = 0$
is certainly equivalent to the conventional lattice gluodynamics.

In the next two sections we study the model (\ref{action}) along the lines $\tilde{\gamma} = 0$
and $\gamma = 0$ in the $(\gamma,\tilde{\gamma})$ parameter space at fixed value of the gauge coupling $\beta$.
The simulations were performed on $20^3$ and $12^4$ lattices at $\beta=6.0$ and $\beta=2.4$ correspondingly.
Note that this choice of parameters is based on the experience with pure YM theory,
in which these $\beta$ values and volumes correspond to the physical scaling regime
\cite{Teper:1998te,Lucini:2001ej}.
While the point $\gamma = \tilde{\gamma}= 0$ was simulated with standard overrelaxed heatbath updating,
away from it we implemented the Metropolis algorithm which is the only one available at non-zero 
$\gamma$, $\tilde{\gamma}$. The procedure turns out to be very time consuming especially in D=4.
Indeed, the one link update step requires to take into account the magnetic charges $q$, $\tilde{q}$
in all neighboring cells the number of which is much larger in D=4 (see Appendix). Because of this we were unable
to thoroughly scan the ample range of $\gamma$-couplings, only the following points were considered
in details
\beq
\label{range}
\!\begin{array}{cccrc}
(\gamma,\tilde{\gamma})_{3D} & = &
        \{(0, 0)\,; & (4, 0)\,, (7, 0)\,, (9, 0)\,; & (0, 4)\}\,, \\
(\gamma,\tilde{\gamma})_{4D} & = &
        \{(0, 0)\,; & (4, 0)\,, (6, 0)\,, (8, 0)\,; & (0, 4)\}\,.
\end{array}
\eeq
In particular, the complexity of the algorithm precludes us from studying the phase diagram
of the model (\ref{action}) (see below) and investigate the finite volume effects.
Below it is silently assumed that the chosen volumes are large enough even at non-zero
$\gamma$, $\tilde\gamma$ couplings.
At each $\gamma$-point we generated about one hundred statistically independent gauge samples
separated by $\sim 10^3$ Monte Carlo sweeps. The observables of primary
importance are the planar Wilson loops from which we extracted the heavy quark potential
(see, e.g. Ref.~\cite{Bali:1994de} for details)
and the correlator of the Polyakov lines $\langle P(0)P(R)\rangle$, $P(\vec{x}) = 1/2 \tr \prod_t U_0(\vec{x}+t)$.
To improve the statistics the standard spatial smearing~\cite{Albanese:1987ds} and hypercubic
blocking~\cite{Hasenfratz:2001hp} for temporal links were used.
In D=4 we also monitored the topological charge $Q$,
the topological susceptibility $\chi = \langle Q^2 \rangle /V$
defined by means of the overlap Dirac
operator~\cite{Neuberger} (see, e.g. Ref~\cite{Giusti:2002sm} for details and further references).

\begin{figure}
\centerline{\psfig{file=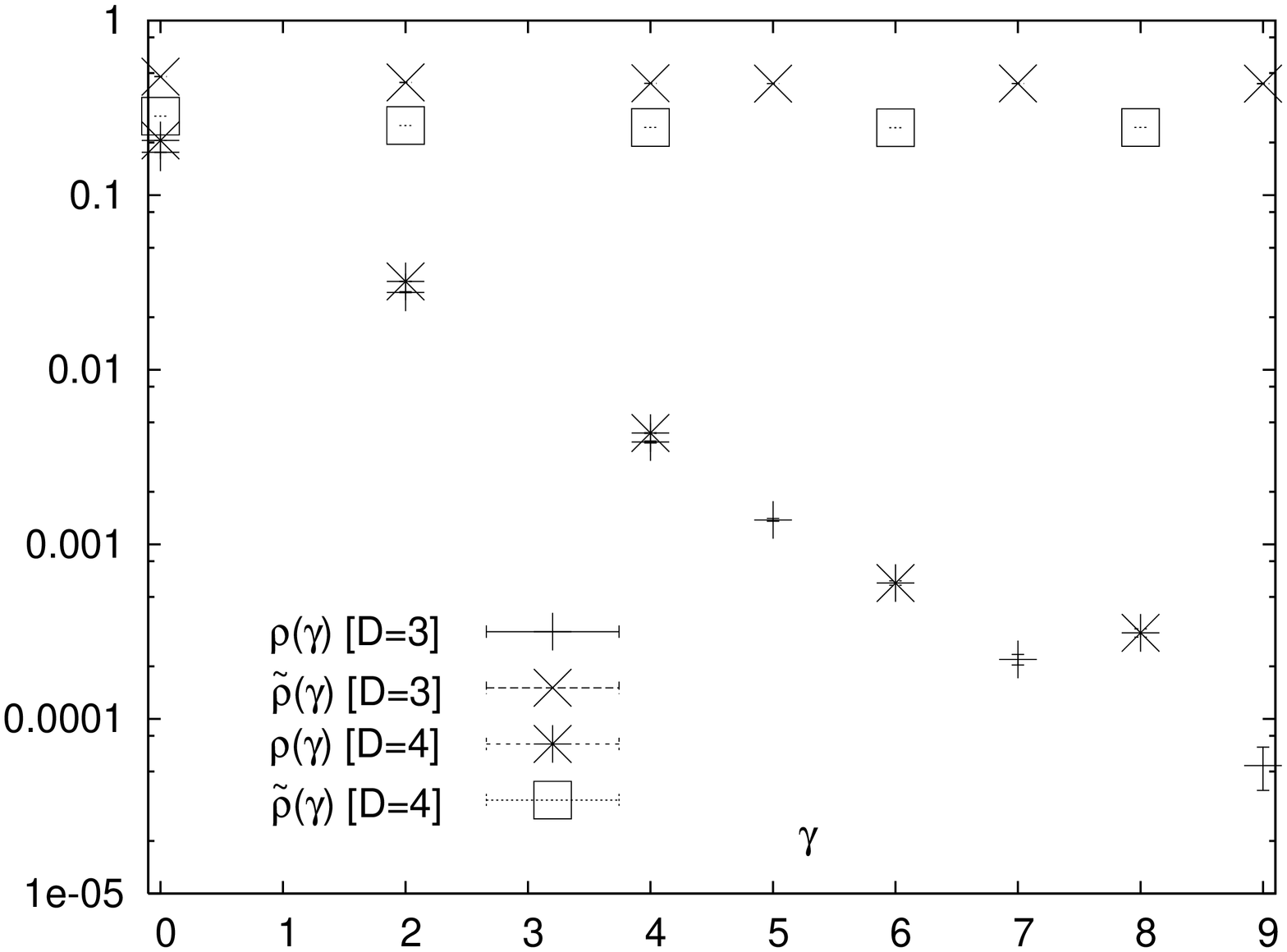,width=0.4\textwidth,silent=,angle=-90}}
\centerline{\psfig{file=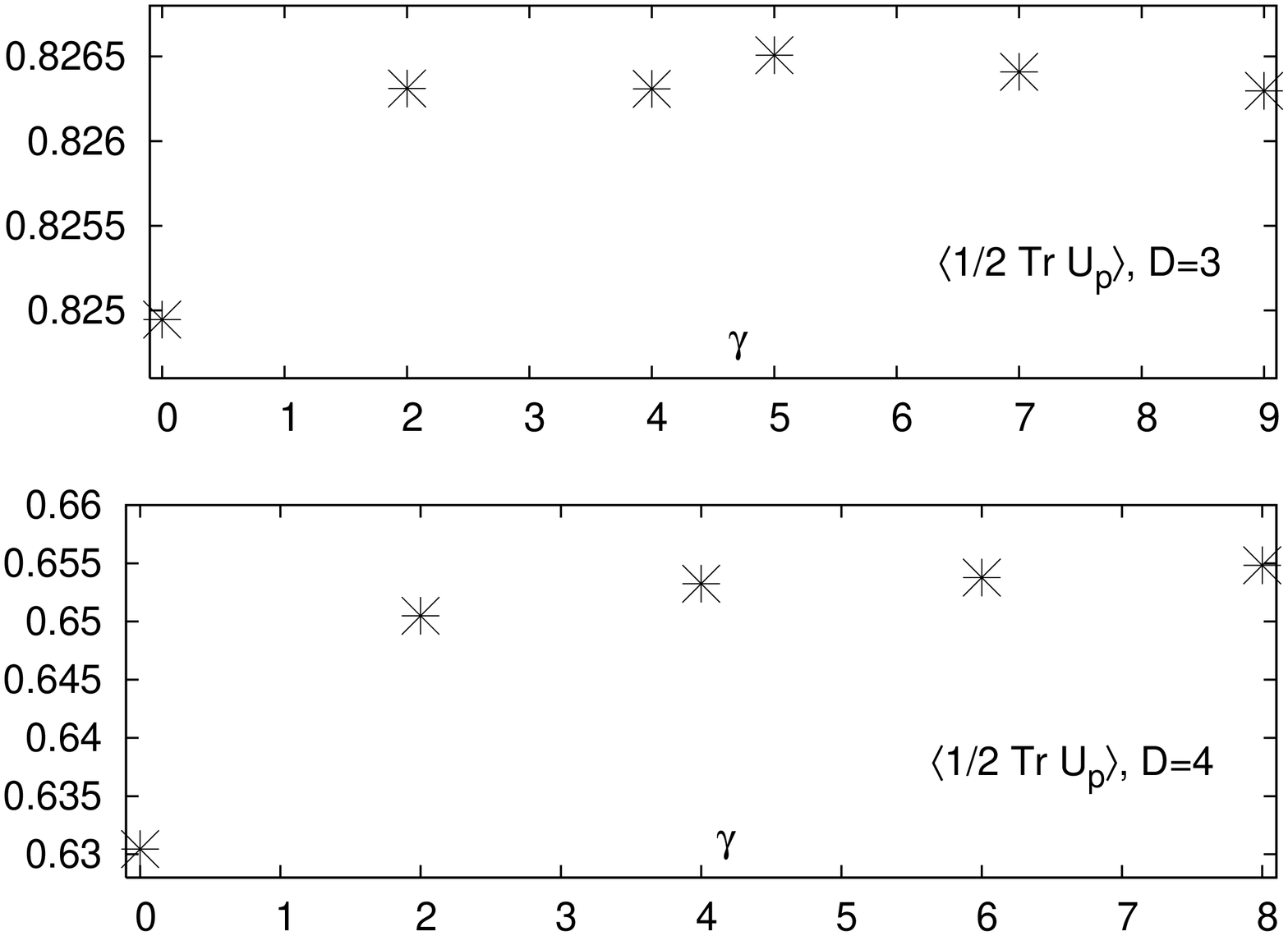,width=0.4\textwidth,silent=,angle=-90}}
\caption{The densities (\ref{rho-cube}), (\ref{rho-site}) and  the mean plaquette $\langle1/2\tr U_p\rangle$
on the line $\tilde{\gamma} = 0$ as functions of $\gamma$.}
\label{fig7}
\end{figure}

\begin{figure}
\centerline{\psfig{file=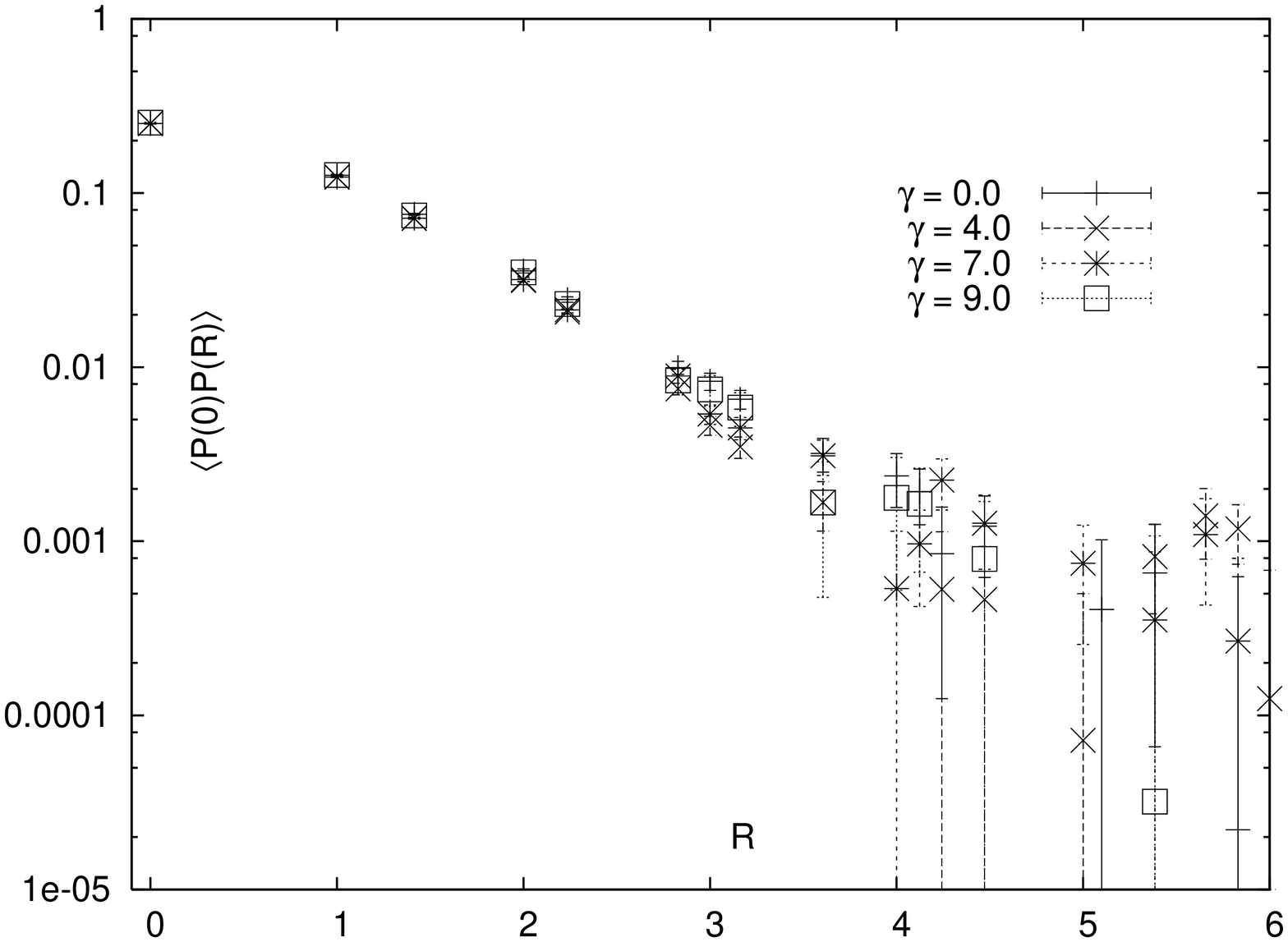,width=0.4\textwidth,silent=,angle=-90}}
\centerline{\psfig{file=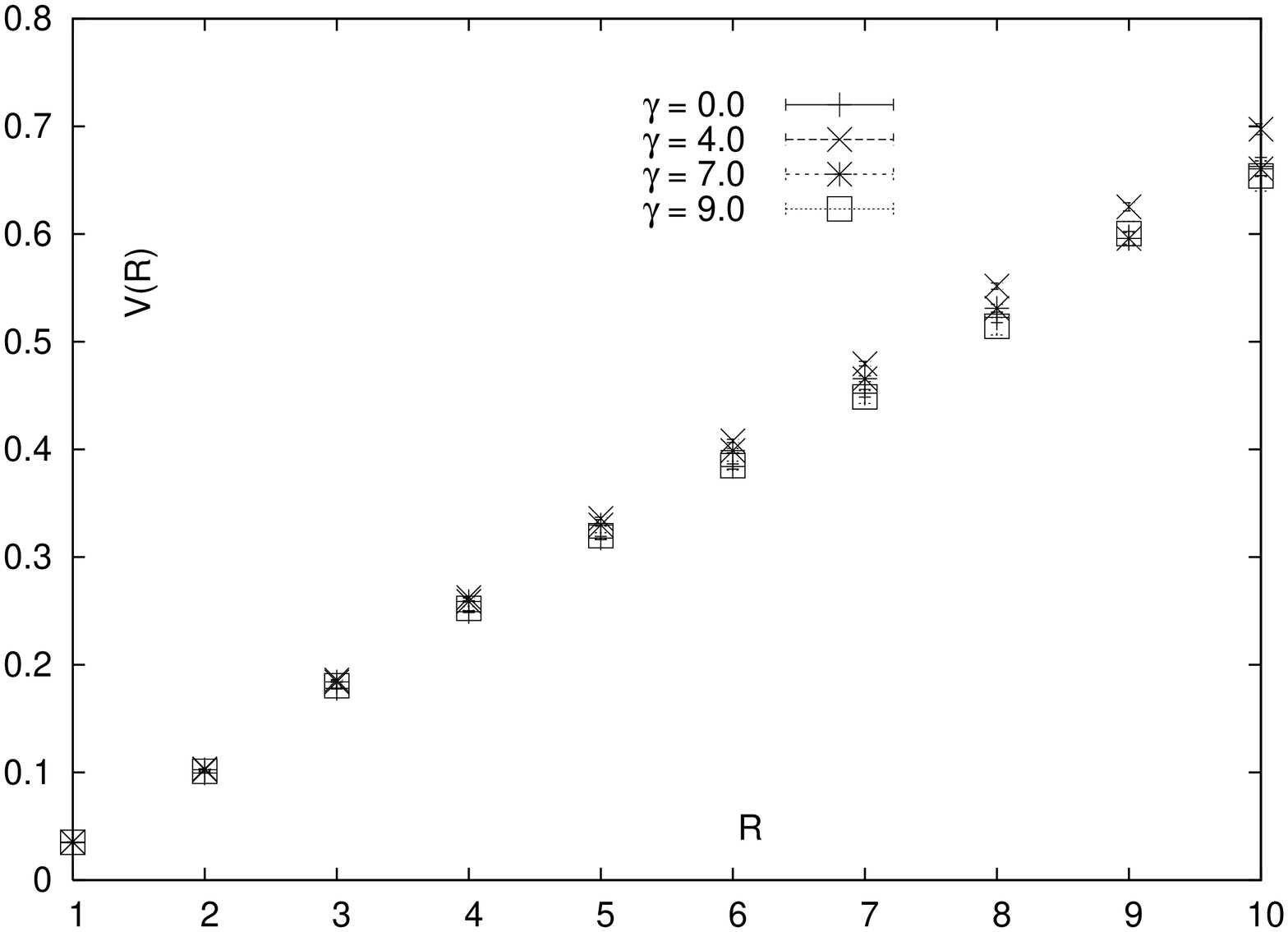,width=0.4\textwidth,silent=,angle=-90}}
\caption{The correlation function of Polyakov lines and the heavy quark potential
in D=3 at $\tilde{\gamma} = 0$ and various $\gamma$.}
\label{fig8}
\end{figure}

\begin{figure}
\centerline{
\psfig{file=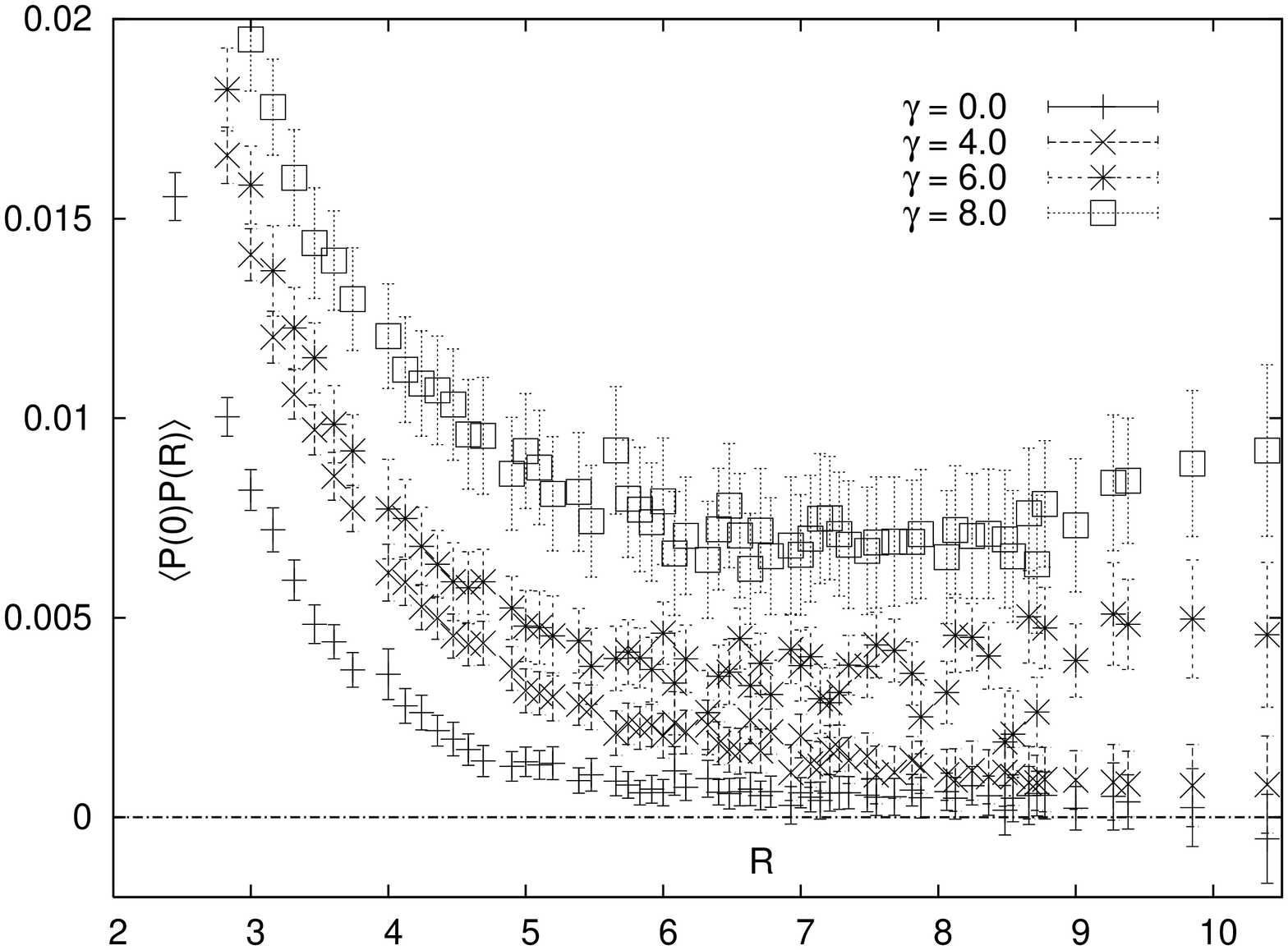,width=0.4\textwidth,silent=,angle=-90}}
\centerline{\psfig{file=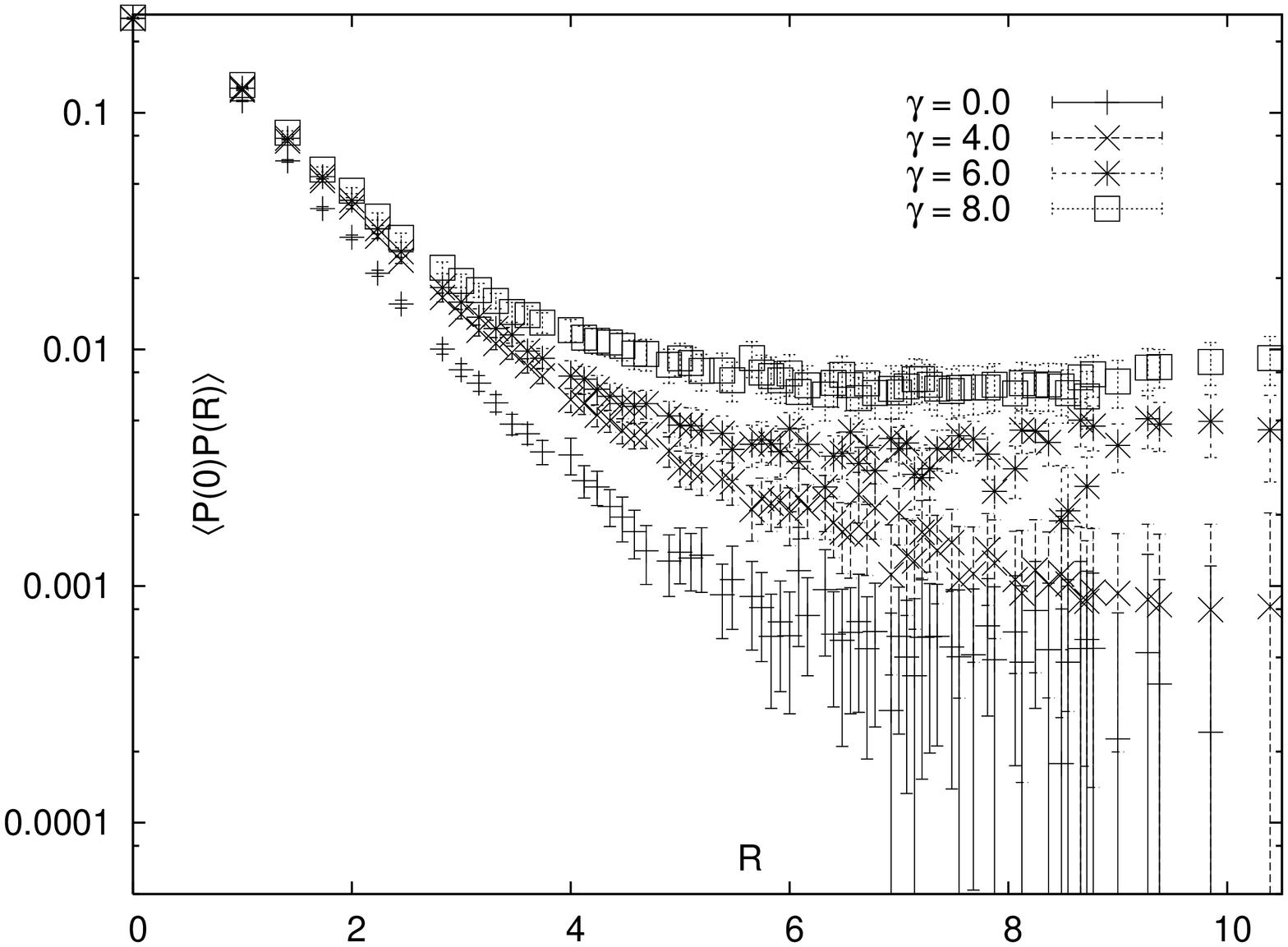,width=0.4\textwidth,silent=,angle=-90}}
\caption{The Polyakov lines correlator in D=4 at $\tilde{\gamma} = 0$ and various $\gamma$.}
\label{fig9}
\end{figure}

\begin{figure}
\centerline{\psfig{file=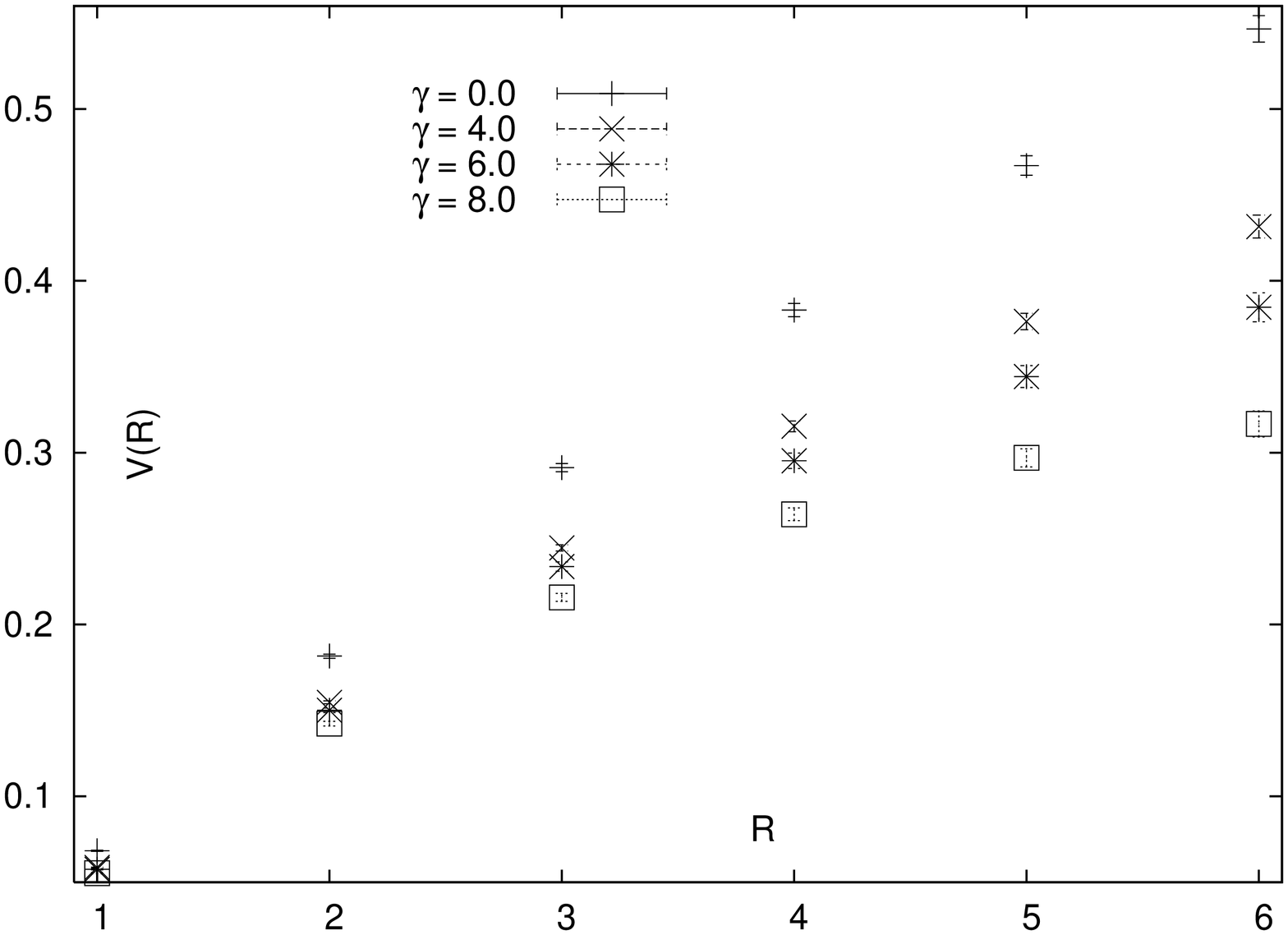,width=0.4\textwidth,silent=,angle=-90}}
\centerline{\psfig{file=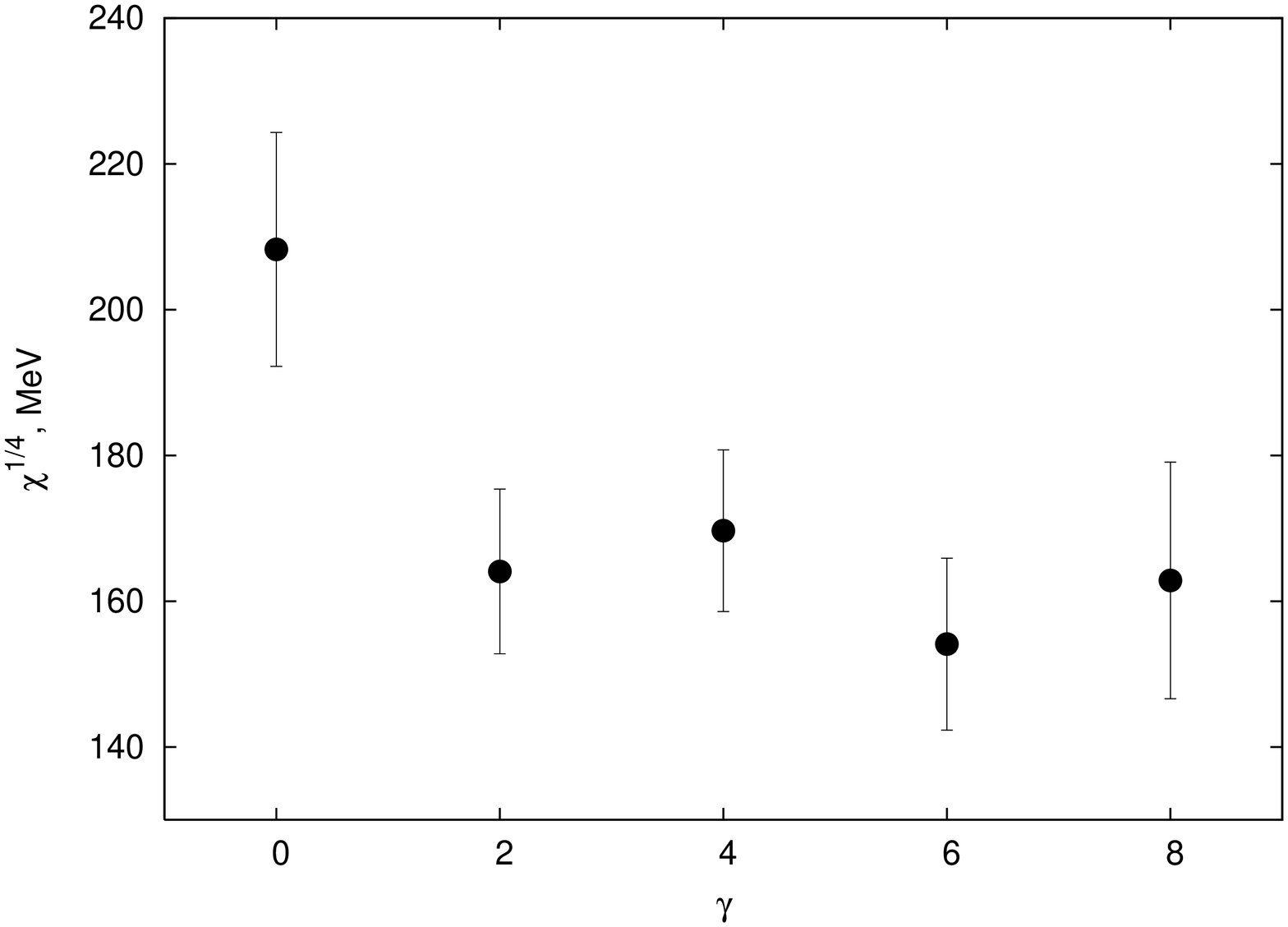,width=0.4\textwidth,silent=,angle=-90}}
\caption{The heavy quark potential $V(R)$ and the topological susceptibility $\chi$
in D=4  at $\tilde{\gamma} = 0$ and various $\gamma$.}
\label{fig10}
\end{figure}

\subsection{$\tilde\gamma = 0$ Line}
\label{gamma_c}
\noindent
Here we study the effect of the gradual removal of the points in which the Bianchi identities
are violated. Let us consider first the behavior of the densities (\ref{rho-cube}), (\ref{rho-site})
with raising $\gamma$ coupling. It turns out that $\tilde{\rho}$ stays almost constant
(Figure~\ref{fig7}, upper panel)
in both three and four dimensions monotonically varying from 0.477(1) to 0.435(1) in D=3
and from 0.2831(1) to 0.2441(2) in D=4 in the entire $\gamma$-range considered.
On the other hand, the density $\rho$ falls down exponentially with $\gamma$ and becomes
of order $O(10^{-4})$ in D=3 ($O(10^{-3})$ in D=4). We note in passing that the mean plaquette
$\langle1/2\tr U_p\rangle$ is also almost insensitive to the $\gamma$ coupling (Figure~\ref{fig7}, bottom)
rising in D=3 from 0.8248(1) to 0.8263(3) when $\gamma$ is changed in the entire range
(the corresponding change in D=4 is from 0.6301(2) to 0.6548(2)).

A few comments are now in order. First, the constancy of $\tilde{\rho}$ and the simultaneous
falloff of $\rho$ by orders of magnitude implies that the above picture of dominating
ultraviolet $q$-$\tilde{q}$ pairs is greatly oversimplified. It seems that the UV fluctuations
are indeed dominating, but their structure is much more involved. In particular, it has little to do
with the model of tightly bounded $q$-$\tilde{q}$ dipoles, rather it is some complicated
mixture of various charge-anticharge configurations which probably don't form the dipole-like pairs at all.

Turn now to the behavior of the heavy quark potential
and the Polyakov lines correlation function with rising $\gamma$ coupling.
In three dimensions (Figure~\ref{fig8}) both the Wilson loops and the correlator
$\langle P(0)P(R) \rangle$ show almost no sign of $\gamma$ coupling dependence,
in particular, the asymptotic string tension at large $\gamma$ is equal to its value
in the pure Yang-Mills theory. However, the situation changes drastically in D=4.
One can see from Figure~\ref{fig9} that the correlation function $\langle P(0)P(R) \rangle$
tends to non-zero positive value at large separations when $\gamma$ coupling becomes of order few units
\beq
\label{PP}
\lim\limits_{R\to\infty} \langle P(0)P(R) \rangle_{\gamma{\scriptscriptstyle\gtrsim} 1} ~=~ const \, > 0\,.
\eeq
The heavy quark potential extracted from Wilson loops is shown on Figure~\ref{fig10} (upper panel) and
for $\gamma\gtrsim 1$ is indeed flattening at large distances
\beq
\lim\limits_{R\to\infty} V_{\gamma{\scriptscriptstyle\gtrsim} 1}(R) ~=~ const \,.
\eeq
Note that it is hardly possible to conclude firmly from Figure~\ref{fig10} along
that the asymptotic string tension is indeed vanishing; however, Eq.~(\ref{PP}) and
Figure~\ref{fig9} are incompatible with its non-zero value.

The other measured observables do not show strong dependence on $\gamma$ coupling.
In particular, the topological charge $Q$ stays at zero in average value albeit with slightly
narrower distribution. As is clear from the bottom panel of Figure~\ref{fig10} the topological
susceptibility $\chi = \langle Q^2 \rangle/V$ diminishes at $\gamma \approx 1$ by approximately 25\%
and the estimation of its limiting value is
\beq
\lim\limits_{\gamma\to\infty} \, \chi^{1/4}(\gamma) ~=~ 163(8)~\mbox{MeV}\,,
\eeq
which should be compared~\cite{Lucini:2001ej} with $\chi^{1/4}(0) = 212(3) \mbox{MeV}$ in pure YM theory,
where the physical units are fixed by the string tension $\sqrt{\sigma}= 440~\mbox{MeV}$.

The discussion of the results presented above is postponed until sec.~\ref{discussions2}.
Here we only note that the dynamics of YM fields in D=3 seems to be almost insensitive
to whether or not the Bianchi identities are violated. In particular, the complete
suppression of the magnetic charges which indicates the violation of the Bianchi identities
has almost no consequences for the correlators we considered. However, the four dimensional
case appears to be quite different. Our results indicate that the suppression of
the Bianchi identities violation is likely to destroy confinement while other measured characteristics
of the theory remain essentially unchanged.

\subsection{Suppressing the Degenerate Points}
\label{gamma_s}
\noindent
Consider the response of the theory on the suppression of the degenerate points.
The qualitative difference in the behavior of the system
along the lines $\gamma=0$ and $\tilde{\gamma}=0$ could be seen
already on the simplest observables like $\rho$, $\tilde{\rho}$.
We have checked that the falloff of the degenerate points fraction
$\tilde{\rho}(\tilde{\gamma})$ is indeed exponential with $\tilde{\gamma}$
in both three and four dimensions; the relevant numbers are
$\tilde{\rho}_{3D}(0) = 0.477(1)$,  $\tilde{\rho}_{3D}(4) = 0.0098(2)$
and 
$\tilde{\rho}_{4D}(0) = 0.2831(1)$,  $\tilde{\rho}_{4D}(4) = 0.045(3)$.
However, the fraction of points at which the Bianchi identities are violated
also notably diminishes with $\tilde{\gamma}$. The falloff of $\rho(\tilde{\gamma})$
in D=3 is not so pronounced ($\rho(0)=0.1758(4)$, $\rho(4)=0.1010(4)$)
and starting from $\tilde{\gamma} \approx 1$ it is numerically larger than $\tilde{\rho}$.
It is surprising, however, that in D=4 the inequality $\rho < \tilde{\rho}$ holds
for all $\tilde{\gamma}$ values considered and in fact the fraction of points at which
the Bianchi identities are violated is diminished by the order of magnitude
(0.2059(2) at $\tilde{\gamma}= 0$ versus 0.024(3) at $\tilde{\gamma}= 4$).
As far as the mean plaquette energy is concerned its behavior is similar to that
on the $\tilde{\gamma} = 0$ line. In particular, in D=3 it essentially stays constant
while in four dimensions it changes from 0.6301(2) to 0.655(1).

As we noted already the suppression of the degenerate points (\ref{q-detB}), (\ref{q-K})
might not be physically meaningful. For instance, in three dimensions
the orientation of the triple $(\vec{B}_1, \vec{B}_2, \vec{B}_3)$,
although being gauge invariant, is not fixed by any symmetry or physical principle.
The attempt to fix the sign of $\mathrm{det}\,B$ everywhere probably will lead to
physically unacceptable results. Indeed, the closer inspection of the Polyakov
lines correlator reveals that it is an oscillating function of the distance.
Hence the lattice reflection positivity is lost and the theory seems to be
pathological at non-zero $\tilde{\gamma}$.

\begin{figure}
\centerline{\psfig{file=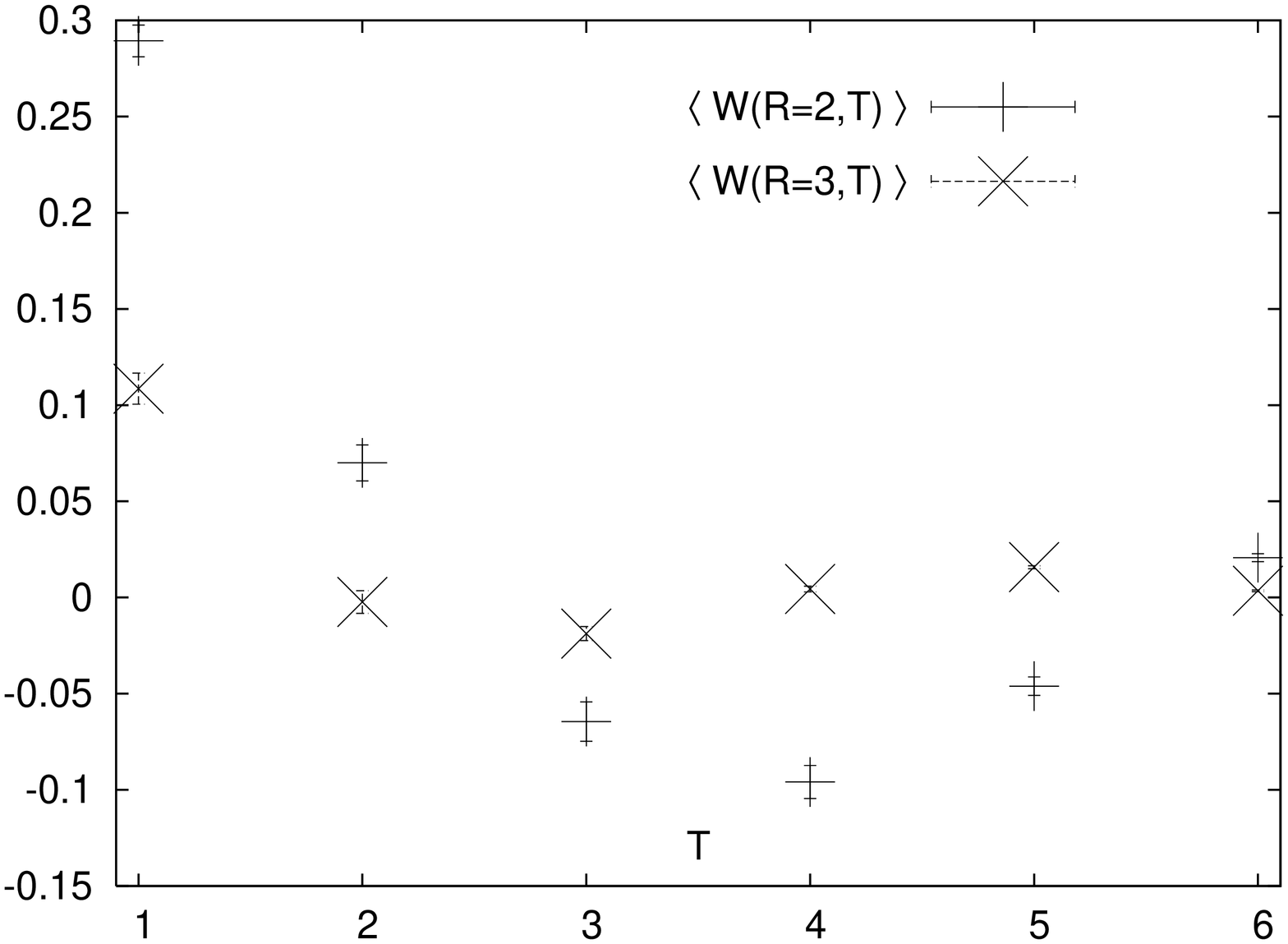,width=0.4\textwidth,silent=,angle=-90}}
\centerline{\psfig{file=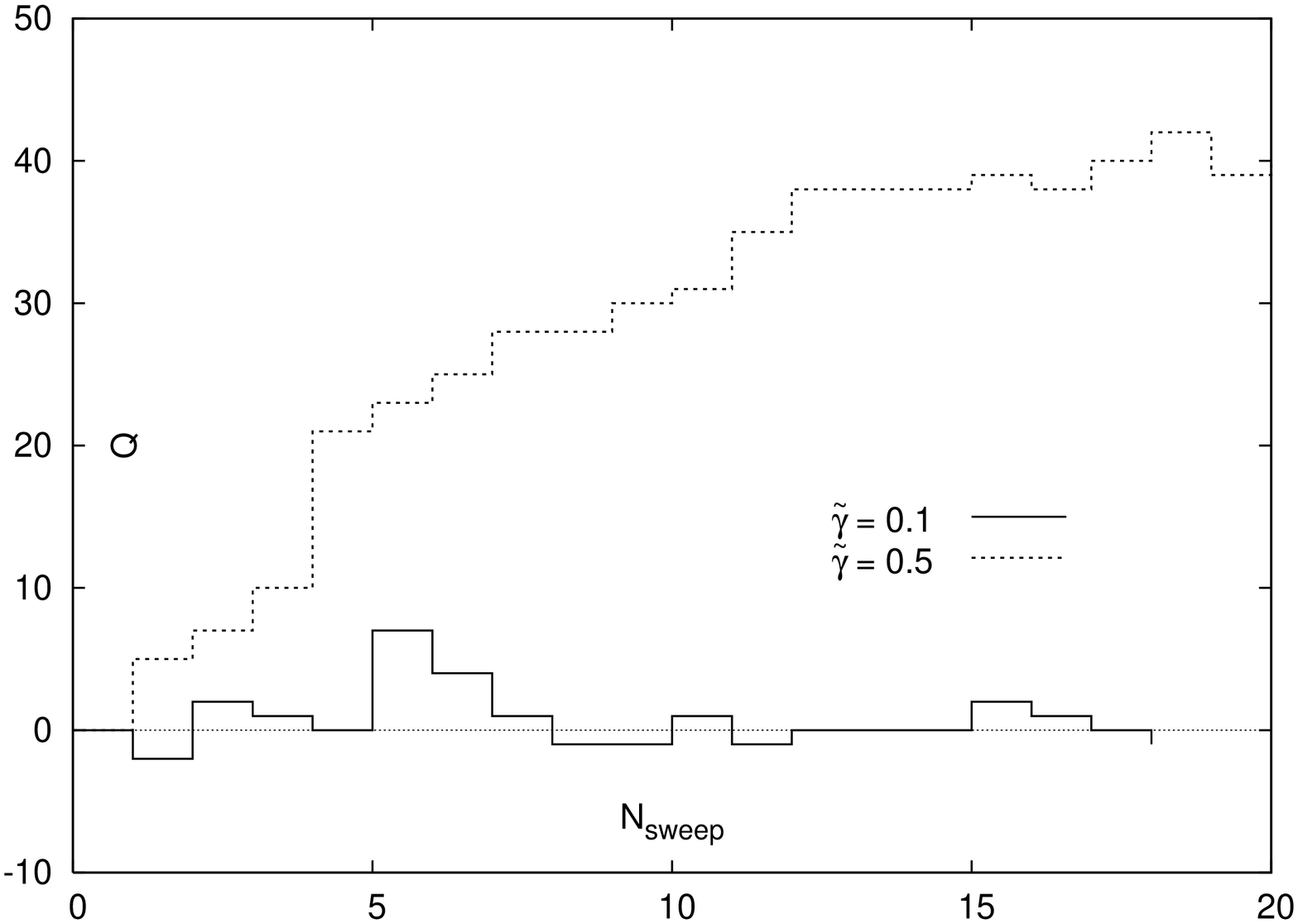,width=0.4\textwidth,silent=,angle=-90}}
\caption{The oscillations of Wilson loops $\langle W(R,T)\rangle$ at $\gamma=0$, $\tilde\gamma=4$ and
the MC history of the topological charge at $\gamma=0$ in D=4.}
\label{fig11}
\end{figure}

In four dimensions the suppression of the degenerate points leads to qualitatively
the same results which, however, are much more pronounced.  For instance, the
Wilson loops $\langle W(R,T) \rangle$ measured at $\tilde{\gamma}\ne 0$
are notably  oscillating at fixed $R$ and varying $T$ (Figure~\ref{fig11}, top panel).
However, unlike the three dimensional case we can easily pin-point the origin
of the reflection positivity violation. Indeed, it is well known that in the fixed
topological sector the theory certainly violates $CP$ and it is natural then to ask 
what is the typical topological charge of the configurations at non-zero $\tilde{\gamma}$.

The bottom panel of the Figure~\ref{fig11} shows the Monte Carlo history of the topological
charge on $8^4$ lattice at $\beta = 2.30$, $\gamma=0$, $\tilde{\gamma}=0.1, 0.5$
when the starting configuration was thermalized at $\gamma=\tilde{\gamma} = 0$
(note that we changed the lattice geometry and the $\beta$ coupling for reasons to be explained shortly).
In view of the observed reflection positivity violation at $\tilde{\gamma}\ne 0$
it is not surprising that $Q$ indeed stays away from zero in average.
What is surprising, however, is that the average topological charge $\langle Q \rangle$
turns out to be always positive and extremely large for $\tilde\gamma > 0$.
In particular, for $0 < \tilde\gamma \ll 1$ the mean topological charge is shifted only slightly
from zero being of order few units. However, once the $\tilde\gamma$ coupling becomes
comparable with unity $Q$ flows away from zero during Monte Carlo updating towards extremely
large positive values with almost constant and very high rate. In fact, it quickly becomes
too large to be technically accessible for us and this was essentially the reason to consider
so small lattices here. The volume dependence of $\langle Q \rangle$ could be inferred
by noting that the last term in (\ref{action}) responsible for the rapid increase of the topological
charge is the bulk quantity. Therefore $\langle Q \rangle$ seems to be proportional
to the volume at fixed $\tilde\gamma$ although we had not thoroughly investigated this dependence numerically.
We have checked that the behavior of $Q$ at non-zero $\tilde{\gamma}$ is always similar
to that on Figure~\ref{fig11}, irrespectively to the concrete meaning
of the last term in the action (\ref{action}).
It does not matter which particular type of magnetic charges $\tilde{q}$ is suppressed by
the $\tilde{\gamma}$ coupling, we always
see the violation of the reflection positivity which is due to the rapid increase
of the global topological charge. This problem is discussed in next section.

\section{Discussions}
\label{discussions2}
\noindent
The interpretation of the results we achieved so far may not be simple and straightforward.
Here we discuss a few particular points which are essential for our work.

First of all, we do see that the physical significance of the Bianchi identities
is quite different in D=3 and D=4. The three dimensional theory turns out to be
insensitive to the suppression of the Bianchi identities violation. Even the complete
removal of $q$ charges from the vacuum does not change the theory in any notable way.
The four dimensional theory seems to be different in this respect.

The suppression of the Bianchi identities violation is likely to destroy confinement liberating
color charges in the fundamental representation. It is tempting to conclude then
that the confinement phenomenon is due to the field configurations for which the
r.h.s. of Eq.~(\ref{bianchi-D4}) is non-vanishing. This conclusion looks natural for the following reasons.
First, it matches the known confinement mechanism in the simple Abelian models.
Secondly, it could explain why in the continuum considerations confinement
is missing since usually the Bianchi identities with
vanishing r.h.s. are taken for granted. Third, it qualitatively
matches the phenomenological lattice observations that the geometrically thin
line- or string-like objects (Abelian monopoles, P-vortices) might be relevant
for confinement (see, e.g Refs.~\cite{Haymaker:1998cw,Greensite:2003bk} for review and further references).
And finally it does not look hopeless from the field-theoretical
point of view since, as we argued above, at vanishing lattice spacing the mechanism
of the Bianchi identities violation has little to do with singular fields, rather it is
related in some complicated way to the points of chromomagnetic fields
degeneracy.

However, the striking difference between three and four dimensional theories 
with respect to the suppression of the Bianchi identities violation shows that this
conclusion is probably misleading. If the confinement phenomenon is indeed due to
the Bianchi identities violation then it should disappear also in D=3 at large $\gamma$ coupling.
But this does not happen and hence we come to the unnatural conclusion that the
confinement mechanism has little in common in D=3 and D=4.

However, we could take a different point of view. Namely, there is indeed a great physical
difference between the Bianchi identities in three and four dimensions. As we discussed
in sec.~\ref{bianchi-intro}, Eq.~(\ref{bianchi-D4}) constitutes
the algebraic restriction on the gauge potentials for a given distribution of the
chromomagnetic fields.
Away from the degeneracy points \footnote{
Note that in the complete axial gauges this reservation is unnecessary.
} the gauge potentials could be completely reconstructed just from the Bianchi identities
along. In this respect the violation of the Bianchi identities could be seen as the source
of the gauge potentials ambiguities and the suppression of non-zero r.h.s.
of Eq.~(\ref{bianchi-D4}) effectively restricts the gauge inequivalent $A^a_\mu$ which
are to be taken into account in the functional integral.
It is crucial that in three dimensions the analogous argumentation fails and in fact
Eq.~(\ref{bianchi-D3}) does not restrict the gauge potentials in any notable
way irrespectively whether or not it is violated.

The natural and probably the only available quantity which is sensitive to the gauge
potentials ambiguities is the $\langle A^2_{min} \rangle$ condensate.
Therefore the following qualitative scenario emerges. It is the non-perturbative
$\langle A^2_{min} \rangle$ condensate which seems to be relevant for confinement.
In four dimensions the Bianchi identities are the tool
which allows to restrict the  $\langle A^2_{min} \rangle$ condensate.
Moreover, the suppression of non-zero r.h.s. of Eq.~(\ref{bianchi-D4})
makes $\langle A^2_{min} \rangle$  to vanish.
Clearly the same approach does not work in three dimensions because the Bianchi identities 
do not constraint $A^a_\mu$ in D=3.
Note that this is only the qualitative picture. In particular, the dependence of
$\langle A^2_{min} \rangle$ on the $q$ charges density could be very complicated
especially because of the dominating perturbative contributions.

To reiterate the point we note that
the justification to consider the modified action (\ref{action})
is that the second term in (\ref{action}) is local and preserves
all the symmetries of the original action. Then the universality suggests that
the continuum limit of the model (\ref{action}) should be $\gamma$ coupling independent
(we take $\tilde\gamma=0$ for definiteness). At the same time our results indicate
that this is probably not the case. If we would accept the Bianchi identities violation
as the primary reason for confinement then we would be faced with serious universality problems.
However, the above scenario based on the $\langle A^2_{min} \rangle$
condensate seems to avoid (at least formally) this issue.

\begin{figure}
\centerline{\psfig{file=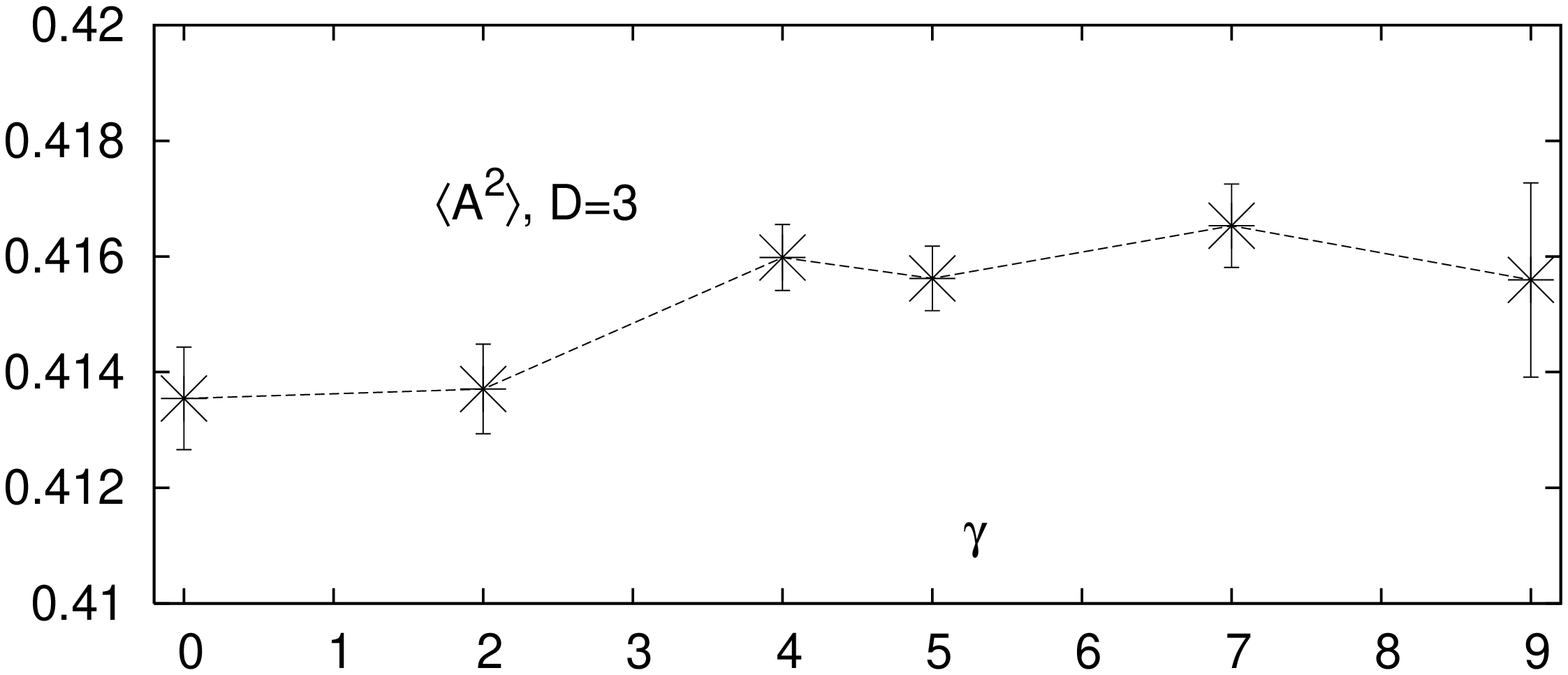,width=0.4\textwidth,silent=,angle=-90}}
\centerline{\psfig{file=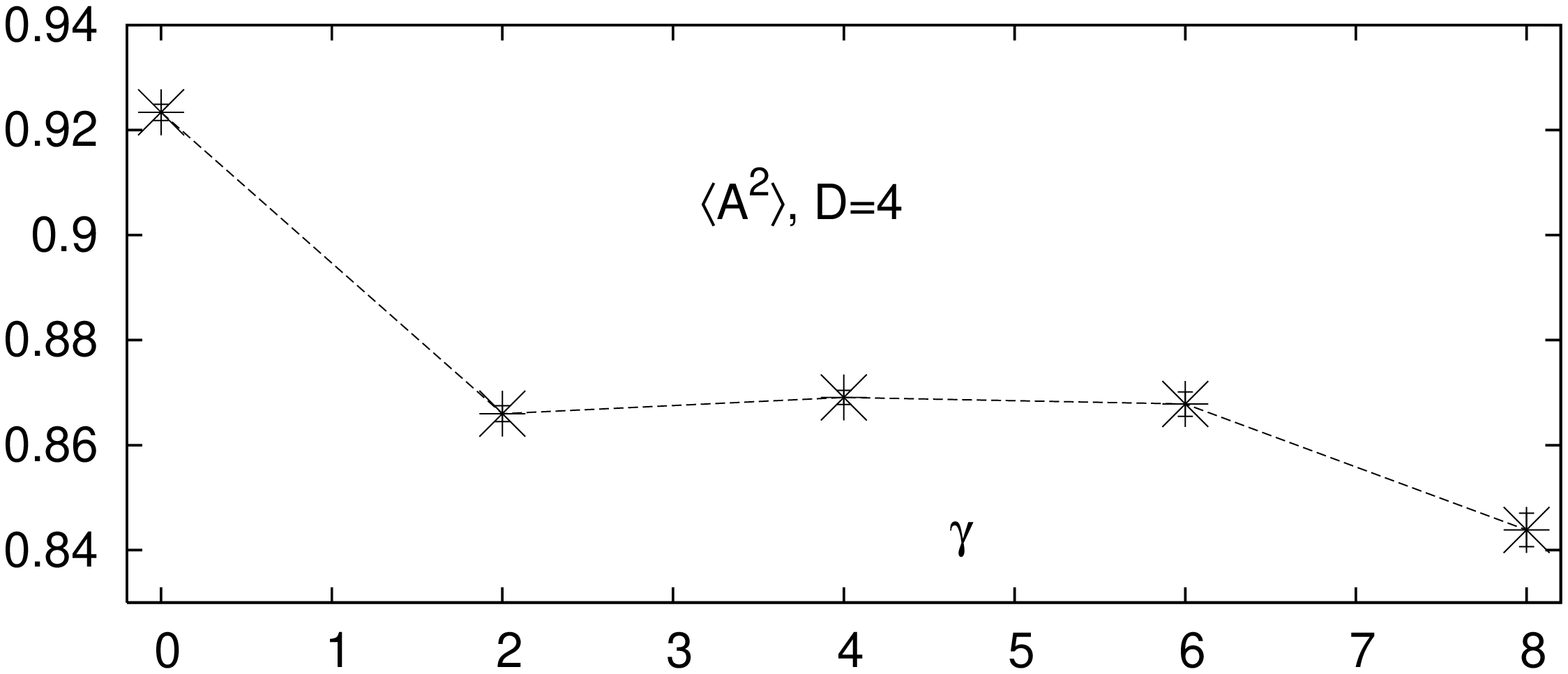,width=0.4\textwidth,silent=,angle=-90}}
\caption{The quantity $\langle A^2 \rangle$ in the Landau gauge at $\tilde\gamma=0$ as function of $\gamma$
in three and four dimensions.}
\label{fig12}
\end{figure}

In fact, the dependence of $\langle A^2_{min} \rangle$ condensate
on the $\gamma$ coupling could be measured directly. Namely, we could measure
the quantity $\langle A^2\rangle$ in the Landau gauge and its drop with rising
$\gamma$ coupling gives an estimate for the behavior of the $\langle A^2_{min} \rangle$
condensate when the violation of the Bianchi identities is gradually removed.
Moreover, this could be compared with the results of Ref.~\cite{Gubarev:2000nz}
where the same Landau gauge $\langle A^2\rangle$ albeit with different normalization
was measured across the finite-temperature deconfinement phase transition.
Note that the quantity $\langle A^2\rangle$ in the Landau and Coulomb gauges
was already introduced in Refs.~\cite{Lavelle:1988eg,Greensite:1985vq}.
The details of our measurements are as follows. The gauge potentials are defined
in terms of the link matrices $U_\mu(x)$
\beq
A^a_\mu ~=~ \tr \, \frac{\sigma^a }{2i\,a}\, [ U_\mu(x) - U^\dagger_\mu(x)]\,,
\eeq
where $a$ is the lattice spacing. The Landau gauge was fixed by minimizing
$\sum_{x,\mu} (A^a_\mu(x))^2$ with overrelaxation algorithm until the magnitude
of $\diff_\mu A^a_\mu$ becomes everywhere less than $10^{-6}$.
The results are presented on Figure~\ref{fig12}.
One can see that in three dimensions $\langle A^2\rangle$ is indeed almost
insensitive to the $\gamma$ coupling confirming the qualitative scenario
outlined above. On the other hand, in four dimensions  $\langle A^2\rangle$
drops down with increasing $\gamma$ by essentially the same amount which was reported
in Ref.~\cite{Gubarev:2000nz}.
Note that the relative drop of $\langle A^2\rangle$ is expected to be small~\cite{Gubarev:2000eu,Gubarev:2000nz}.
Indeed, on general grounds we have
$$
\langle A^2\rangle  = \frac{1}{a^2}\,\left(\,\sum\limits_n b_n \alpha^n_s + a^2 \langle A^2_{min}\rangle\,\right)
$$
and clearly the Landau gauge $\langle A^2\rangle$ is dominated by the perturbative tail at weak coupling.
However, the drop in $\langle A^2\rangle$ across the phase transition is believed to be entirely due
to the non-perturbative condensate $\langle A^2_{min}\rangle$.

The next comment concerns the behavior of the topological charge with respect to the
degenerate points suppression. It is true that the violation of the reflection positivity
with rising $\tilde\gamma$ coupling is to be expected on the general
grounds. Moreover, it is also expected that in D=4 the non-zero in average global topological
charge is the origin of the reflection positivity violation. However, the following
questions remain: why the topological charge in always positive and 
rises so rapidly with respect to the Monte Carlo updating? What is the relation between $Q$
and the magnetic charges $q$, $\tilde{q}$?

Evidently, the non-zero in average value of the topological charge requires it to be mostly either
positive or negative. At the same time our experience with various possible definitions
of the last term in the action (\ref{action}) shows that the positivity of $Q$ at 
$\tilde\gamma \ne 0$ is seemingly built in our approach from very beginning. 
As far as we can see the only
place which distinguishes between $Q \gtrless 0$ is the canonical orientation of the
elementary plaquettes which we accepted (see sec.~\ref{chromomagnetic-fields}).
Indeed, the construction of the magnetic charges $q$, $\tilde q$ depends on the particular canonical
orientation which is not uniquely defined. Although there are only few possibilities
to choose from, different choices could discriminate the sign of the topological charge.
Indeed, in the fermionic language the sign of $Q$ distinguishes the left and right chiralities
(orientations) analogously to the canonical orientation which discriminates 
the left and right coordinate systems.
Thus we expect that the sign of $Q$ at non-zero $\tilde\gamma$ will change with inequivalent
choice of the canonical orientation.
As far as the rapid growth of $Q$ is concerned it seems that the only possible explanation
is that the non-zero $\tilde\gamma$ coupling lifts the degeneracy of different topological
sectors. The situation is reminiscent to the quantum mechanical problem of the periodic
potential on which the constant $\tilde\gamma$-dependent electric field is superimposed.

It would be instructive to have the explicit expression for $Q$ in terms of the $q$, $\tilde q$
magnetic charges distribution. However, the usual approaches available in the
literature (see, e.g. Refs.~\cite{Reinhardt:1997rm,Jahn:1998nw})
seemingly lead to the erroneous results. For instance, the treatment of Ref.~\cite{Jahn:1998nw}
applies almost literally in our case. The outcome is that the topological charge is
given by the linear combination of the $q$, $\tilde q$ magnetic charges
and hence vanishes when $q$, $\tilde q$ are highly suppressed. This seems to contradict
the observed rapid growth of $Q$ with rising $\gamma,\tilde\gamma$ couplings when all the
$q$, $\tilde q$ charges are suppressed on equal footing.
Therefore either the results of Ref.~\cite{Jahn:1998nw} should be modified in our case or we should look for
different definition of the topological charge.
The definition of the topological charge, which closely follows the approach
of the present work, could be given along the lines of Refs.~\cite{Avron:1988ty,Johnsson:1997ub}
(see also Ref.~\cite{Demler:1998pm} for an excellent introduction).
Instead of studying the evolution of single spinor along the closed contours we could consider
the corresponding evolution of the degenerate two-level system for which the resulting
non-Abelian geometrical phase describes the YM instanton. This approach is under
investigation and will be published elsewhere.

\section{Conclusions}
\noindent
In this paper we considered the non-Abelian Bianchi identities in SU(2) pure
Yang-Mills theory focusing on the physical significance of the chromomagnetic fields
degeneracy points and the possibility of Bianchi identities violation. These questions
necessitate the regularization and we specifically kept in mind the lattice formulation.
It had been known for a long time that the Bianchi identities in general are the requirement
that the gauge holonomy for any null-homotopic path equals to unity. The main achievement
of this paper is the reformulation of the above requirement in terms of the physical elementary
fluxes (field strength). Our approach is based on the non-Abelian Stokes theorem
appeared recently and allows to give an explicit gauge invariant expression for the Bianchi
identities on the lattice. Simultaneously it allows to formulate the notion of the non-Abelian
Bianchi identities violation in gauge invariant and local form.

As a further development of our approach we showed that the study of the lattice Bianchi identities
naturally leads to the consideration of the chromomagnetic fields degeneracy points
at which a particular determinants constructed from $E^a_i$, $B^a_i$ vanish.
It turns out that the violation of the Bianchi identities and the degenerate points are closely
related to each other. In particular, in the weak coupling regime the Bianchi identities violation
is not related generically to the singular fields, rather it is due to the existence
of the degenerate points.

As is clear from the above presentation the main advantage of our approach is that the non-Abelian
nature of the theory had been traded for the complicated geometry which, however, allows
the pure geometrical Abelian-like treatment. Then both the Bianchi identities violation
and the degeneracy points formally appear as usual magnetic charges. However, we stress
that the term ``magnetic charge'' and, in fact, the entire Abelian analogy is only formal.
In particular, the physical interpretation of $q$, $\tilde{q}$  charges is completely different;
there is no magnetic charge conservation whatsoever on the original (hyper)cubical lattice.
Nevertheless, the Abelian-like representation is invaluable for the analysis presented above.

The locality and gauge invariance of the definition of the Bianchi identities and
the chromomagnetic fields degeneracy points permits us to modify the original gauge action and to study the effects
of gradual removal of these objects from the vacuum. It turns out that in the four dimensional
case the suppression of the Bianchi identities violation seems to be relevant for confinement: the heavy
quark potential extracted from Wilson loops flattens at large distances and the correlator
of the Polyakov lines tends to non-zero constant at large separations.
At least this is the case on the lattices we have studied.
At the same time, other correlation functions which we measured had not been changed considerably.
The situation in D=3 turns out to be just opposite. Namely, the theory is almost insensitive
to the suppression of the Bianchi identities violation.
However, in D=4 the complexity of the numerical simulations precluded us from studying
the relevant issues like the phase diagram of the modified model, the volume dependence
of our results, etc. We hope to address these questions elsewhere.

As far as the degenerate points are concerned any attempt to remove them from the vacuum
results in the reflection positivity violation. Moreover, in D=4 this violation is due to the
extremely large positive global topological charge which grows rapidly during Monte Carlo
updating. This observation could be relevant for studying the gluodynamics in the topologically
non-trivial sectors.

Confronting the results obtained in D=3,4 we argued that it is probably misleading to consider
the violation of the Bianchi identities as the primary cause of confinement. Instead the correct
picture would be to interpret the Bianchi identities as an algebraic constraint on the gauge potentials and
to relate the confinement phenomenon to the existence of the non-perturbative
$\langle A^2_{min} \rangle$ condensate. This scenario seems to be in agreement with universality
expectations, works the same in both three and four dimensions and does not contradict our findings.

\begin{acknowledgments}
\noindent
The authors are grateful to prof. V.I.~Zakharov, E.T.~Akhmedov and to the members
of ITEP lattice group for stimulating discussions. This work was partially supported
by grants RFBR-03-02-16941, RFBR-05-02-16306a, RFBR-05-02-17642 and RFBR-0402-16079.
The work of F.V.G. was supported by INTAS YS grant 04-83-3943.
\end{acknowledgments}

\appendix
\section*{Appendix}
\noindent
Here we describe the cell complex  underlying the Bianchi identities
(\ref{bianchi-lattice}). We start from single plaquette and note that
the application of non-Abelian Stokes theorem (\ref{NAST-lattice})
assigns spinor wave function $\bra{z}$ to each plaquette corner.
This could be represented by 4 points belonging to
this plaquette and shifted from the corners towards the plaquette center.
The totality of this points constitutes the 0-skeleton $\mathbb{C}^0$ of the
cell complex and it is convenient to parametrize $s\in \mathbb{C}^0$ by the
point $x$ of the original lattice and by two shifts
with corresponding shift directions (see Figure~\ref{fig5}, left):
\bea{c}
s = s(x,\,\mu,d_\mu,\,\nu,d_\nu\,) = s(x,\,\nu,d_\nu,\,\mu,d_\mu\,)\,, \\ \\
\mu \ne \nu\,, \qquad d_\mu, d_\nu = \pm 1\,.
\eea
In total there are $2 D(D-1)\cdot V$ sites, where $V$ is the lattice volume.

Turn now to the 1-skeleton $\mathbb{C}^1$ which consists of two types of links.
The first group contains the original links
\beq
\label{app-link1}
~\!\!\!\!\!\! s_i = s(x,\mu,d_\mu,\nu,d_\nu) \to s_f = s(x+\hat{\mu},\mu,-d_\mu,\nu,d_\nu),
\eeq
which carry the matrix element $\bra{z(s_i)}\,U_\mu(x)\,\ket{z(s_f)}$. Links from the second group
\bea{ccc}
\label{app-link2}
s_i = s(x,\mu,d_\mu,\nu,d_\nu) & \to & s_f = s(x,\lambda,d_\lambda,\nu,d_\nu) \\
& \mu \ne \lambda & \\
\eea
are ascribed with the matrix element $\braket{z(s_i)}{z(s_f)}$.

As far as the 2-skeleton $\mathbb{C}^2$ is concerned its structure is different 
in three and four dimensions. As a consequence $\mathbb{C}^k$, $k > 2$, also
differ considerably and are described separately below.

\subsection*{D=3}
\noindent
Here $\mathbb{C}^2$ contains three types of 2-cells. First, there are original plaquettes $p$
the boundary of which consists of the links (\ref{app-link1}).
Moreover, the standard coboundary operator $d: \mathbb{C}^1\to\mathbb{C}^2$
acts in accordance with Eq.~(\ref{total-flux}) and assigns the flux
magnitude $\Phi(p)$ to the plaquette. 
Second group of 2-cells  contains various squares $\mathscr{S}^{(1)}(x,\mu,d_\mu)$
constructed from links (\ref{app-link2}). Namely, $\mathscr{S}^{(1)}(x,\mu,d_\mu)$
has the following four points in its boundary
\beqn
\mathscr{S}^{(1)}(x,\mu,d_\mu): &
s(x,\,\mu,d_\mu,\,\nu,\pm d_\nu)\,, &
s(x,\,\mu,d_\mu,\,\lambda,\pm d_\lambda)\,, \nonumber \\
\label{app-square1}
& \mu \ne \nu \ne \lambda\,. &
\eeqn
The coboundary operator $d: \mathbb{C}^1\to\mathbb{C}^2$ assigns the Bargmann
invariant (\ref{omega-x}) to the particular square. The argumentation of sec.~\ref{bianchi}
allows to show that
\beq
\label{app-square}
\mathscr{S}^{(1)}(x,\mu,1) ~=~ \mathscr{S}^{(1)}(x+\hat{\mu},\mu, -1 )\,,
\eeq
where we have denoted the values assigned to each square by the same symbol ``$\mathscr{S}^{(1)}$''
hoping that this will not lead to confusion. Third type of 2-cells contains
various triangles $\mathscr{T}^{(1)}$ constructed from links (\ref{app-link2});
there are three points in the boundary of $\mathscr{T}^{(1)}$
\beqn
\label{app-triangle}
\mathscr{T}^{(1)}: & s(x,\,\mu,d_\mu,\,\nu,d_\nu)\,, \\
& s(x,\,\nu,d_\nu,\,\lambda,d_\lambda)\,, \,\, s(x,\,\lambda,d_\lambda,\,\mu,d_\mu)\,, \nonumber \\
& \mu \ne \nu \ne \lambda\,. \nonumber
\eeqn
The operator $d: \mathbb{C}^1\to\mathbb{C}^2$
assigns the corresponding Bargmann invariant to the triangle.
Note that the last group of 2-cells is formed by mixture of links (\ref{app-link1}), (\ref{app-link2})
and need not be considered, in fact: by Eq.~(\ref{basis-change}) the phase associated with them
is always zero. It is important that the value assigned by $d$ to every 2-cell
is always taken modulo $2\pi$ and is rather similar to $\theta_{plaq} = [d\theta]_{2\pi}$
in the language of compact U(1) gauge model. In other words it is silently assumed that
only gauge invariant quantities are ascribed to every 2-cell.

As far as the 3-skeleton $\mathbb{C}^3$ is concerned it contains essentially two types
of 3-cells. First, the original lattice cubes which
look as on Figure~\ref{fig4} (right); each cube contains 6 plaquettes and 8 triangles
at its corners. The coboundary operator $d: \mathbb{C}^2\to\mathbb{C}^3$ considered for any particular cube
is identical to Eq.~(\ref{bianchi-lattice}) by construction.
The 3-cells of the second group are constructed entirely from triangles $\mathscr{T}^{(1)}$ and squares
$\mathscr{S}^{(1)}$ above and are illustrated on Figure~\ref{fig5} (right). 
The physical meaning of the corresponding magnetic charge is analyzed in sec.~\ref{detB}.

Thus the consideration of three dimensional case is completed. Note that geometrically
there is one more type of 3-cells which, however, need not be taken into account.
These 3-cells are formed by two squares (\ref{app-square1}) and four links (\ref{app-link1})
connecting them. It follows from (\ref{basis-change}) and (\ref{app-square}) that
$d: \mathbb{C}^2\to\mathbb{C}^3$ always gives zero on these cells.

\subsection*{D=4}
\noindent
In four dimensions the consideration of the cell complex underlying the lattice Bianchi identities
(\ref{bianchi-lattice}) becomes cumbersome.
In particular, we do not give the full list of cells forming $\mathbb{C}^k$, $k=2,3,4$,
only cells relevant to the considerations in sec.~\ref{detB}, \ref{numerics}
are presented.

First we note that the D=3 construction applies directly in D=4. In particular, 2-skeleton
includes the plaquettes, squares (\ref{app-square1}) and triangles (\ref{app-triangle}),
trivially generalized to four dimensions. In the 3-skeleton $\mathbb{C}^3$ we identify
then the usual 3-cubes and 3-cells shown on Figure~\ref{fig5} (right).

\begin{figure}
\centerline{\psfig{file=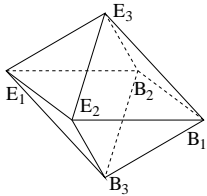,width=0.15\textwidth,clip=,silent=}}
\caption{$\mathscr{D}^{(2)}$ 3-cell in D=4 (see the text).}
\label{app-fig}
\end{figure}

However, it is clear that in D=4 the $\mathbb{C}^2$, $\mathbb{C}^3$ are not exhausted
by the above 2- and 3-cells. In particular, the 2-skeleton contains now an additional set
of triangles $\mathscr{T}^{(2)}$ with vertices
\beqn
\label{app-triangle-D4}
\mathscr{T}^{(2)}:     &   s(x,\mu,d_\mu,\nu,d_\nu)\,, \\
  & s(x,\mu,d_\mu,\lambda,d_\lambda)\,,\,\, s(x,\mu,d_\mu,\rho,d_\rho)\,, \nonumber \\
  &  \mu \ne \nu \ne \lambda \ne \rho\,, \nonumber
\eeqn
and squares $\mathscr{S}^{(2)}$, the vertices of which are
\beqn
\label{app-square-D4}
\mathscr{S}^{(2)}: & s(x,\mu,d_\mu,\nu,d_\nu)\,,\,\, s(x,\nu,d_\nu,\lambda,d_\lambda)\,, \\
                   & s(x,\lambda,d_\lambda,\rho,d_\rho)\,,\,\,  s(x,\rho,d_\rho,\mu,d_\mu)\,, \nonumber \\
                   & \mu \ne \nu \ne \lambda \ne \rho\,. \nonumber
\eeqn
All these 2-cells are constructed from links (\ref{app-link2}) and therefore are ascribed with
the appropriate Bargmann invariants.

In the 3-skeleton $\mathbb{C}^3$ the new diamond-like cells $\mathscr{D}$
consisting of 6 vertices and 8 triangles appear. In turn, these 3-cells could be
subdivided into two groups.

$\mathscr{D}^{(1)}(x,\mu,d_\mu)$: the 3-cells in this group are constructed
from 8 triangles (\ref{app-triangle-D4})
and are similar to those considered in D=3, Figure~\ref{fig5} (right).
In particular, one can show that (cf. Eq.~(\ref{app-square}))
\beq
\mathscr{D}^{(1)}(x,\mu, 1) ~=~ \mathscr{D}^{(1)}(x+\hat{\mu},\mu, -1)\,,
\eeq
see the note following Eq.~(\ref{app-square}). The physical interpretation of the corresponding
magnetic charge is discussed in sec.~\ref{detB}.

$\mathscr{D}^{(2)}(x,d_\mu,d_\nu,d_\lambda,d_\rho)$: these 3-cells are built from both types of triangles
(\ref{app-triangle}), (\ref{app-triangle-D4}). The corresponding vertices are constructed
by fixing a particular combination of shift directions $d_\mu$: there are 6 distinct planes passing
through given lattice site in D=4 and $s(x,\mu, d_\mu,\nu,d_\nu)$, $\mu\ne\nu$ is one of the six vertices
of $\mathscr{D}^{(2)}(x,d_\mu,d_\nu,d_\lambda,d_\rho)$ cell. The total number of these 3-cells 
per lattice site is $2^4 = 16$.
Note the specific pattern of the flux directions assigned to the vertices of
$\mathscr{D}^{(2)}(x, d_\mu,d_\nu,d_\lambda,d_\rho)$,
which is radically different from what we have encountered so far. In the weak coupling limit
the opposite vertices are ascribed with the same components of chromoelectric and chromomagnetic
fields (see the Figure~\ref{app-fig}). 


\end{document}